\documentclass[12pt]{article}
\usepackage{setspace}
\usepackage{amsfonts}
\usepackage{mathrsfs}
\usepackage{dsfont}
\usepackage{amsmath}
\usepackage{multirow}
\usepackage{graphicx}
\usepackage{amssymb}
\usepackage{subfig}
\usepackage{mathtools}
\usepackage{bm}
\usepackage{natbib}
\usepackage{bbm}
\usepackage{hyperref}
\usepackage{verbatim}
\usepackage{lscape}
\usepackage{titlesec}
\usepackage[affil-it]{authblk}
\usepackage{pifont}
\usepackage{array, multirow, bigdelim, makecell, booktabs} 
\usepackage[dvipsnames]{xcolor}

\usepackage{enumitem}
\usepackage[paperwidth=9in,tmargin=0.7in,lmargin=1in, rmargin=1in]{geometry}
\renewcommand{\abstractname}{Abstract}
\makeatletter
\renewenvironment{abstract}{%
    \if@twocolumn
      \section*{\abstractname}%
    \else 
      \begin{center}%
        {\bfseries \Large\abstractname\vspace{\z@}}
      \end{center}%
      \quotation
    \fi}
    {\if@twocolumn\else\endquotation\fi}
\makeatother

\begin{document}
\title{Using auxiliary marginal distributions in imputations for nonresponse while accounting for survey weights, with application to estimating voter turnout}
\author{Jiurui Tang, D. Sunshine Hillygus, Jerome P. Reiter}
\date{}
\maketitle

\begin{abstract}
\normalsize
\noindent 
The Current Population Survey is the gold-standard data source for studying who turns out to vote in elections.  However, it suffers from potentially nonignorable unit and item nonresponse.  Fortunately, after elections, the total number of voters is known from administrative sources and can be used to adjust for potential nonresponse bias.
We present a model-based approach to utilize this known voter turnout rate, as well as other population marginal distributions on demographic variables, in multiple imputation for unit and item nonresponse.  In doing so, we ensure that the imputations produce design-based estimates that are plausible given the known margins. We introduce and utilize a hybrid missingness model comprising a pattern mixture model for unit nonresponse and selection models for item nonresponse. Using simulation studies, we illustrate repeated sampling performance of the model under different assumptions about the missingness mechansisms.  
We apply the model to examine voter turnout by subgroups using the 2018 Current Population Survey for North Carolina.  As a sensitivity analysis, we examine how results change when we allow for over-reporting, i.e., individuals self-reporting that they voted when in fact they did not.
\end{abstract}


\section{Introduction}

When data suffer from unit nonresponse (no values are observed for some units) or item nonresponse (some values are observed and some values are missing for some units), analysts generally need to make strong assumptions about the reasons for missingness.  For example, they may need to assume data are missing at random \citep{Rubin1976, Mealli_Rubin_2015} or that they follow a specific nonignorable missingness mechanism \citep{linero:daniels}.  Such assumptions are inescapable with missing data.

One way to reduce reliance on strong assumptions is to utilize auxiliary information from external data sources.  Here, we consider sets of population percentages or totals for categorical variables. For example, we may have accurate estimates of population percentages of demographic variables from censuses, large-sample surveys, or administrative databases.  To motivate how such auxiliary information can help with missing data, suppose we have data from a survey that includes a question on whether or not one votes in an election.  The survey suffers from missingness on vote, which we want to (multiply) impute to facilitate estimation of the turnout rate. Among survey respondents, 70\% indicate that they voted in the election.   However, using population size estimates and post-election administrative data, we know that only 50\% of voters turned out.
Assuming respondents do not misreport, this auxiliary marginal information suggests that the missingness for vote is nonignorable; that is, people who did not vote are more likely not to tell their voting status. Knowing the margin, we can impute values for the missing votes so that the completed data estimates of turnout are closer to 50\% than 70\%.  Without this margin, we would likely impute the missing votes using a missing at random (MAR) assumption---which would perpetuate the nonresponse bias---or have to make some heroic, unverifiable assumption about the vote distribution  of the nonrespondents.

In this article, we present methods for leveraging auxiliary marginal distributions with application to estimating turnout in the 2018 U.S.\ Congressional election in North Carolina using the 2018 Current Population Survey (CPS).  To do so, we build on the missing data with auxiliary margins (MD-AM) framework  developed by \citet{Akande2021MDAM}, which we review  in Section \ref{MDAM}.  Specifically, because the CPS relies on a complex sampling design---as do many large-scale government and social surveys---we extend the MD-AM framework to account for survey weights from unequal probability sampling designs. 
%
The basic idea is to generate multiple imputations so that the completed data result in plausible design-based estimates of the known margins.  This strategy was introduced by  \citet{AkandeWeights}, who illustrated it for stratified simple random samples subject to item nonresponse only.  We leverage and extend this strategy to surveys with weights generated by other complex sampling designs and subject to both item and unit nonresponse. 
In doing so, we introduce a hybrid missing data mechanism that uses a pattern mixture model formulation for unit nonresponse and a selection model formulation for item nonresponse.  Using simulation studies, we show that using a hybrid missingness MD-AM model can result in more reasonable survey-weighted inferences than approaches that do not utilize the margins.  We then use a hybrid missingness MD-AM model to estimate voter turnout for subgroups in the 2018 CPS.  More broadly, the CPS analysis is illustrative of a general, multiple imputation strategy for using auxiliary margins to handle nonignorable missing data in surveys with complex sampling designs.




The remainder of this article is organized as follows. In Section \ref{MDAM}, we review the MD-AM framework of \cite{Akande2021MDAM} for simple random samples, explaining how it can leverage auxiliary margins to handle unit and item  nonresponse.  
In Section \ref{MDAM.new}, we present an MD-AM model that uses the hybrid missingness mechanism and accounts for survey weights. We also describe methods for estimating the parameters of this model and generating multiple imputations of missing values, and we  
summarize results of simulation studies evaluating the model fitting procedure.  The simulation studies are described in detail in the supplementary material.
In Section \ref{CPS}, we use a hybrid missingness MD-AM model to estimate turnout in demographic subgroups using the 2018 CPS in North Carolina.  We also assess the sensitivity of results to potential reporting error in the vote responses.  In Section \ref{conclusion}, we provide some concluding remarks.


\section{MD-AM Modeling with Simple Random Samples}
\label{MDAM}

When presenting the MD-AM framework for simple random samples, \citet{Akande2021MDAM} describe a two step process.  In step 1, the analyst specifies a joint distribution for the survey variables and for the indicators for nonresponse that are identifiable from the observed data alone.  This model typically  uses default assumptions about the missingness, such as MAR or itemwise conditionally independent nonresponse \cite[ICIN, ][]{Sadinle_Reiter2017}.  In step 2, the analyst, informed by the available margins, adds parameters to the model while ensuring that the model as a whole can be identified as in \cite{Sadinle2019}.  Adding these parameters weakens the assumptions about the missingness mechanisms. 
In this section, we review the MD-AM framework for simple random samples, 
beginning with general notation that we use throughout.  

\subsection{Notation}
\label{notation}
Let $\mathcal{D}$ comprise the data that were planned to be collected from a survey of $i=1,\dots,n$ individuals, and let $\mathcal{A}$ comprise the auxiliary margins. Let $X = (X_1,\dots,X_k)$ represent the $p$ variables in both $\mathcal{A}$ and $\mathcal{D}$, where each $X_k = (X_{1k},\dots,X_{nk})^T$ for $k=1,\dots,p$. Let $Y = (Y_1,\dots,Y_q)$ represent the $q$ variables in $\mathcal{D}$ but not in $\mathcal{A}$, where each $Y_k = (Y_{1k},\dots,Y_{nk})^T$ for $k = 1,\dots,q$. We disregard variables in $\mathcal{A}$ but not in $\mathcal{D}$ as they are not of primary interest. We assume that $\mathcal{A}$  contains only sets of marginal distributions for variables in $X$, e.g., available in some external database. 

For each $k=1, \dots, p$, let $R_{ik}^x=1$ if individual $i$ would not respond to the question on $X_k$ in the survey and $R_{ik}^x=0$ otherwise. Similarly, for each $k=1,\dots,q$, let $R_{ik}^y=1$ if individual $i$ would not respond to the question on $Y_k$ in the survey and $R_{ik}^y=0$ otherwise. Let $R^x = (R_1^x,\dots,R_p^x)$ and $R^y = (R_1^y, \dots, R_q^y)$ be the vectors of item nonresponse indicators for variables in $X$ and $Y$ respectively, where each $R_k^x = (R_{ik}^x,\dots,R_{nk}^x)^T$ and $R_k^y = (R_{ik}^y,\dots,R_{nk}^y)^T$. 
Let $U_i=1$ when individual $i$ in $\mathcal{D}$ does not respond to the survey, so that we do not have their responses. Let $U_i = 0$ when observe at least some values of the survey variables for individual $i$. Let $U = (U_1, \dots, U_n)$. 

\subsection{Illustrative Model Specification}
\label{model}

We illustrate the ideas underpinning  the  MD-AM framework using a  $\mathcal{D}$ comprising two binary variables, $X_1$ and $X_2$.  We have auxiliary information for both from $\mathcal{A}$; thus, this illustrative example does not include any $Y$ variables (whereas our 2018 CPS analysis does use $Y$ variables). We suppose $X_1$ suffers from item nonresponse and $X_2$ is fully observed, so that $R_1^x$ is the vector of item nonresponse indicators for $X_1$. For units with $U_i = 1$, we do not observe $(X_{1i}, X_{2i}, R_{1i}^x)$.   We do not concern ourselves with $R_2^x$ since $X_2$ is fully observed for cases with $U_i=0$.  $U$ is fully observed for all records in $\mathcal{D}$. Figure \ref{data} displays the observed data and the auxiliary information on $X_1$ and $X_2$.  We use this same example to introduce the hybrid missingness MD-AM model in Section \ref{MDAM.new}. 

\begin{figure}[t]
\centering
\caption[Illustrative data for MD-AM.]{Illustrative scenario for the MD-AM model in simple random samples.  $\mathcal{D}$ comprises two binary variables, $X_1$ and $X_2$. $X_2$ is fully observed; $X_1$ suffers from item nonresponse; and, some units do not respond. Here, ``$\checkmark$" represents observed values, ``?" represents missing values.}
\begin{tabular}{|c |c |c |c |l}
\cline{1-4}
$X_1$ & $X_2$ & $R_1^x$ & $U$\\
\cline{1-4}
$\checkmark$ & $\checkmark$ & 0 &\multirow{2}{*}{0} & \hspace{-0.2em}\rdelim\}{3}{1mm}[$\mathcal{D}$] \\
\cline{1-3}
? & $\checkmark$ & 1 & \\
\cline{1-4}
? & ? & ? & 1 \\
\cline{1-4}
$\checkmark$ &  &  &  & \hspace{-0.2em}\rdelim\}{2}{1mm}[$\mathcal{A}$]\\
\cline{1-4}
 & $\checkmark$ &  &  \\
\cline{1-4}
\end{tabular}
\label{data}
\end{figure}

\citet{Akande2021MDAM} recommend using selection models for all missingness indicators.  With their modeling strategy, we write the joint distribution of $(X_1,X_2,R_1^x,U)$ as a product of sequential  conditional distributions, using selection models for the unit and item nonresponse indicators. 
\begin{align}
\mathbb{P}(X_1, X_2,R_1^x,U) &= \mathbb{P}(R_1^x | X_1, X_2, U) \mathbb{P}(U | X_1, X_2)  \mathbb{P}(X_1|X_2) \mathbb{P}(X_2).\label{jointdistr}
\end{align}
The joint distribution in \eqref{jointdistr} can be fully parameterized with fifteen parameters plus the constraint that probabilities sum to one. However, $\mathcal{D}$ alone provides enough information to identify only six parameters. This is evident in the six observable quantities in Figure \ref{data}, namely the probabilities for $U$ and $(R_1^x \mid U=0)$, the two probabilities for $(X_2 \mid R_1^x=r, U=0)$ where $r \in \{0,1\}$, and the two probabilities for $(X_1 \mid X_2=x, R_1^x=0, U=0)$ where $x \in \{0,1\}$. Thus, we need to make some identifying assumptions about the missingness in order to estimate the joint distribution.

The auxiliary information about the marginal distributions of $X_1$ and $X_2$ add constraints on the joint distribution in \eqref{jointdistr}. The MD-AM framework leverages this  additional information.  We begin with step 1: specify an identifiable model based on $\mathcal{D}$ alone with at most six parameters. 
Using selection models for the nonresponse indicators, one natural choice is
\begin{align}
X_{2} &\sim Bernoulli(\pi_{x_{2}}),\,\,\, logit(\pi_{x_{2}}) = \alpha_0 \label{sel_x2}\\
X_{1}|X_{2} &\sim Bernoulli(\pi_{x_{1}}),\,\,\, logit(\pi_{x_{1}}) = \beta_0 + \beta_1 X_{2} \label{sel_x1}\\
U |X_1, X_2 &\sim Bernoulli(\nu_0) \label{sel_unit}\\
R_1^x |U, X_1, X_2 &\sim Bernoulli(\pi_R),\,\,\, logit(\pi_{R}) = \gamma_0 + \gamma_1 X_2 \label{sel_Rx}.
\end{align}
This default model has unit nonresponse following a missing completely at random mechanism (MCAR) mechanism and item nonresponse in $X_1$ following an ICIN mechanism. 

In step 2, we add one term related to $X_1$ and one term related to $X_2$ to the specifications in \eqref{sel_unit} -- \eqref{sel_Rx}. One approach is to add both terms to the model for unit nonresponse in \eqref{sel_unit}, leaving the model for item nonresponse as in \eqref{sel_Rx}.  With slight re-use and abuse of notation, we replace \eqref{sel_unit} with 
\begin{equation}
U \mid X_1, X_2 \sim Bernoulli(\pi_{U}),\, logit(\pi_{U}) = \nu_0 + \nu_1 X_1 + \nu_2 X_2. 
\label{sel_x2_u}
\end{equation}
This is an example of an additive nonignorable model \citep{Hirano2001, Deng2013, Schifeling:Reiter,yajuan:reiter:hilly} for unit nonresponse.  It includes the MCAR mechanism as a special case ($\nu_1=\nu_2=0$), but allows for nonignorable mechanisms as well (either $\nu_1\neq0$ or $\nu_2\neq0$). Thus, the data can inform whether MCAR is plausible, or if not allow for unit nonresponse to depend on $X_1$ or $X_2$.  

There is a key identifying assumption in \eqref{sel_x2_u}: there is no interaction between $X_1$ and $X_2$.  With only marginal information on $X_1$ and $X_2$, we have no way to identify, for example, a differential effect of $X_1$ on $U$ at $X_2=1$ versus $X_2=0$.  If we had the joint margin of $(X_1, X_2)$, we could add their interaction to \eqref{sel_x2_u}.  Nonetheless, using $\mathcal{A}$ has allowed us to weaken substantially the strong MCAR assumption in \eqref{sel_unit}. 

As an alternative to \eqref{sel_x2_u}, we could specify an additive nonignorable model for $R^x_1$, which would allow the mechanism for item nonresponse in $X_1$ to depend on the value of $X_1$ itself.  In doing so, however, we have to remove the direct dependence of $U$ on $X_1$, as the margin for $X_1$ only provides enough information to estimate one additional parameter. Thus, again re-using notation for the regression parameters, we could replace \eqref{sel_unit} and \eqref{sel_Rx} with  the alternative specification,
\begin{align}
U |X_1, X_2 &\sim Bernoulli(\pi_U),\,\,\, logit(\pi_{U}) = \nu_0 + \nu_1 X_2 \label{sel_unita}\\
R_1^x |X_1, X_2, U &\sim Bernoulli(\pi_R),\,\,\, logit(\pi_{R}) = \gamma_0 + \gamma_1 X_1 + \gamma_2 X_2.
\label{sel_set2.4a}
\end{align}

As this example suggests, analysts can use $\mathcal{A}$ to enrich the unit nonresponse model or item nonresponse model. 
\citet{Akande2021MDAM} suggest that analysts use the auxiliary margins to help model unit nonresponse when the unit nonresponse rate is higher than the item nonresponse rate or when the primary concern is about nonignorable unit nonresponse.  Similarly, analysts can enrich the item nonresponse model when item nonresponse is a bigger concern than unit nonresponse.
Researchers can fit multiple model specifications and compare their results as a sensitivity analysis.  

In the CPS analysis described in Section \ref{CPS}, unit nonresponse is more substantial than item nonresponse.  Hence, when describing our adaptation of the MD-AM framework for handling unequal survey weights, we focus on using $\mathcal{A}$ to enrich modeling of unit nonresponse.

\section{MD-AM with Hybrid Missingness Model and  Weights}\label{MDAM.new}



The MD-AM model in Section \ref{MDAM} is intended for simple random samples.  In complex designs, it can result in biased estimates of finite population quantities, as the modeling and imputation steps do not account for unequal probabilities of selection.  \citet{AkandeWeights} describe how to modify the MD-AM model to account for the special case of stratified simple random sampling subject to item nonresponse only.  To do so, they use a rejection sampler: one proposes imputations of the missing values, computes design-based estimates of totals for variables with known margins, and accepts or rejects the proposals according to how close in distribution the estimated totals are the known marginal total. Their model employs selection models for the item nonresponse indicators.

We attempted to apply the selection modeling approach in \eqref{jointdistr} and the strategy in \citet{AkandeWeights} to account for more general unequal probability sampling designs with both item and unit nonresponse.  However, 
the acceptance rate of the rejection sampler is generally too low to be practically useful, and it can be difficult to generate imputations that result in plausible design-based estimates for arbitrary survey weights.  We instead propose a hybrid missingness model that uses a pattern mixture model for unit nonresponse and a selection model for item nonresponse. Using the example in Section \ref{MDAM}, we specify a model for $U$, a model for $(X_1, X_2 \mid U)$, and a model for $(R_1^x \mid X_1, X_2, U)$, as we now describe.  

We modify step 1 in the MD-AM framework to reflect the alternate factorization.  Again re-using some notation for regression parameters, we have the default model,  
\begin{align}
U &\sim Bernoulli(\pi_U) \label{mix_unit}\\
X_{2}|U &\sim Bernoulli(\pi_{x_{2}}),\,\,\, logit(\pi_{x_{2}}) = \alpha_0 \label{mix_x2}\\
X_{1}|X_{2},U &\sim Bernoulli(\pi_{x_{1}}),\,\,\, logit(\pi_{x_{1}}) = \beta_0 + \beta_1 X_{2} \label{mix_x1}\\
R_1^x|X_1, X_2, U &= Bernoulli(\pi_{x_{1}}),\,\,\, logit(\pi_{R}) = \gamma_0 + \gamma_1 X_2 \label{mix_Rx}.
\end{align}
This is essentially the same model proposed for step 1 in Section \ref{model}, re-organized here to clearly show the pattern mixture modeling for $U$.  
We have to assume $R_1^x$ is conditionally independent of $U$ given $\mathcal{D}$ to enable identification of $(\gamma_0,\gamma_1)$, as we have no information to distinguish distributions of $R_1^x$ for unit respondents and nonrespondents.  


Following step 2 of the MD-AM framework, we next add terms related to $X_1$ and $X_2$ to leverage the information in the known margins. We dedicate both terms to unit nonresponse modeling by adding an indicator variable for unit nonresponse, which we label as $U$, to \eqref{mix_x2} and \eqref{mix_x1}.  We have 
\begin{align}
X_{2}|U &\sim Bernoulli(\pi_{x_{2}}),\,\,\, logit(\pi_{x_{2}}) = \alpha_0 + \alpha_1 U \label{mix_x2_u}\\
X_{1}|X_{2},U &\sim Bernoulli(\pi_{x_{1}}),\,\,\, logit(\pi_{x_{1}}) = \beta_0 + \beta_1 X_{2} + \beta_2 U. \label{mix_x1_u}
\end{align}
Here, \eqref{mix_x2_u} implies that the distribution of $X_2$ differs for unit nonrespondents and unit respondents, and $\alpha_1$ controls the strength of nonresponse bias. Specifically, $logit(\pi_{x_2}) =  \alpha_0$ for unit respondents, and $logit(\pi_{x_{2}}) = \alpha_0 + \alpha_1$ for unit nonrespondents. As a result, with larger $|\alpha_1|$, the distribution of $X_2$ is less homogeneous across unit nonrespondents and unit respondents. Similar interpretations can be applied for \eqref{mix_x1_u}.  
Essentially, 
we can view \eqref{mix_x1_u} as a model with different intercepts for unit respondents and unit nonrespondents, but with slopes preserved.  

For the hybrid missingness MD-AM model, one key identifying assumption is no interactions with $U$.  This is unavoidable when using only univariate margins.  With bivariate margins, analysts can add interaction effects. For example, when the joint margin of $(X_1, X_2)$ is known, analysts can add an interaction between $U$ and $X_2$ in \eqref{mix_x1_u}.

\subsection{Incorporating Survey Weights}
\label{weight_UNa}

We now incorporate survey weights into the hybrid missingness MD-AM model.  Our overarching goal is to facilitate valid design-based estimation, which is the preferred approach to finite population inference for many practitioners. Let $N$ denote the population size from which the $n$ survey units in $\mathcal{D}$ are sampled. For $i=1, \dots, N$, let $\pi_i$ be the probability that  unit $i$ is selected to be in the sample, and let $w_i^d = 1/\pi_i$ be its design weight (also known as its sampling weight). 


\subsubsection{Using Margins for Probabilistic Constraints}
\label{weight_UN}

In finite population inference, often we seek to estimate the population totals or means of survey variables. For example, we may seek to estimate the total of some variable $X_k$, which we write as 
$T_{X_k} = \sum_{i=1}^NX_{ik}$.
Analysts can estimate $T_{X_k}$ with the \citet{HT1952} estimator, 
\begin{equation}
\hat{T}_{X_k} = \sum_{i \in \mathcal{D}}w_i^d X_{ik}.
\end{equation}
Finite population central limit theorems ensure that when $n$ is large enough, for fully observed data we have 
\begin{equation}
\hat{T}_{X_k} \sim N(T_{X_k},V_{X_k})
\label{prob_cons_origin}
\end{equation}
with some variance $V_{X_k}$.  Analysts can estimate $V_{X_k}$ using design-based principles \citep{Fuller_2009}.

 However, we cannot compute $\hat{T}_{X_k}$ directly when units in $\mathcal{D}$ are missing values of $X_k$, either due to unit or item nonresponse.  Nonetheless, \eqref{prob_cons_origin} still holds for the unobserved value of $\hat{T}_{X_k}$. Furthermore, we know $T_{X_k}$ (or the population percentage, $T_{X_k}/N$) for any $X_k$ with a margin in $\mathcal{A}$.
 Thus, as suggested in \citet{AkandeWeights}, we should impute the missing values for $X_k$ so  that any completed data set produces a reasonable $\hat{T}_{X_k}$ based on \eqref{prob_cons_origin}. 
 
 We implement this probabilistic constraint in the illustrative model with binary $X_1$ and $X_2$ as follows. For all $i \in \mathcal{D}$ and for all $k$, let $X_{ik}^\star = X_{ik}$ when $R_{ik}^x = 0$, and let $X_{ik}^\star$ be an imputed value when $R_{ik}^x = 1$ or $U_i=1$. In addition to following \eqref{mix_Rx} -- \eqref{mix_x1_u}, the imputations of missing $(X_1, X_2)$  should satisfy, 
\begin{equation}
\sum_{i \in \mathcal{D}}w_i^d X_{i1}^\star \sim N(T_{X_1},V_{X_1}),\,\,\,\,\,\,\,\,\sum_{i \in \mathcal{D}}w_i^d X_{i2}^\star \sim N(T_{X_2},V_{X_2}).\\
\label{prob_cons}
\end{equation}

When values of $X_k$ are missing, the design-based estimate of $V_{X_k}$ also is not directly computable.  
We therefore consider $V_{X_k}$ as a parameter set by the analyst to reflect how closely $\hat{T}_{X_k}$ should match $T_{X_k}$ in any completed data set. When we set $V_{X_k}$ to be relatively small, we encourage imputations that result in $\hat{T}_{X_k}$ close to $T_{X_k}$.  We note, however, that with  small $V_{X_k}$ it can be challenging computationally  to find imputations that satisfy \eqref{prob_cons} and adequately explore the space of plausible imputations of the missing values \citep{TangDis}.
On the other hand, when we let $V_{X_k} \to \infty$, we effectively impose no probabilistic constraint on $\hat{T}_{X_k}$, in which case the MD-AM model is not identifiable.  

We suggest setting $V_{X_k}$ to be plausibly close to what its design-based estimate would have been absent missing data.  To do so, analysts can use the following procedure. 
First, generate a completed data set by imputing $X_{ik}$ for units with $R_{ik}^x=1$ using a common MAR mechanism, such as via multiple imputation by chained equations \citep{MICE2011} or, if available, by using the imputations made by the agency responsible for collecting $\mathcal{D}$. Second, compute the survey-weighted estimate of variance of $\hat{T}_{X_k}$ with this completed data set. We denote this estimate as   $\widetilde{V}_{X_k}$ and use it for $V_{X_k}$.  For the survey-weighted estimation, we assume that the analyst uses respondent weights that are adjusted for unit nonresponse, for example, by the agency responsible for collecting $\mathcal{D}$.  %


\subsubsection{Weights for Unit Nonrespondents}\label{weightsunit}
Data analysts do not always have access to design weights for survey nonrespondents; indeed, federal agencies routinely do not provide any weights for survey nonrespondents.  Therefore, we develop methods for scenarios where weights are unavailable for unit nonrespondents.  For convenience in notation, when using design weights in estimates, we set $w_i^d = 0$ whenever $U_i = 1$. 


The procedure for estimating the MD-AM model, which we describe in section \ref{MCMC}, requires the availability of weights for all units. When the design weights for survey respondents are available as part of $\mathcal{D}$, we need to create weights for the survey  nonrespondents. 
To do so, we treat all nonrespondents equally and let each have the same weight.  We generate weights for nonrespondents so that the sum of the weights for all $n$ units in $\mathcal{D}$ equals $N$. 
 Therefore, when design weights are available, the weights that we use for analysis are as follows:
\begin{equation}
w_i =  \left\{\begin{array}{lr}
        w_i^d & \text{if } U_i = 0\\
        \frac{N-\sum_{j \in \mathcal{D}} w_j^d}{\sum_{j \in \mathcal{D}} U_j} & \text{if } U_i = 1.
        \end{array}\right. \label{known_weights_unit_res}
\end{equation}

\subsubsection{Using Adjusted Weights}
\label{sec:adjwts}


It is a common practice for design weights to be adjusted for unit nonresponse so that the respondents weight up to the full population \citep{Valliant2013}. Indeed, adjusted weights, which we denote as $w_i^a$, are often the only weights available to analysts. Weight adjustment methods include but are not limited to weighting class adjustment, propensity score adjustment, poststratification and raking  \citep{Valliant2013,Lohr2010,yajuanBayesraking}. If we  use adjusted weights in \ref{known_weights_unit_res}, 
we may end up with very small or even negative values of $w_i$ because the sum of the $w_i^a$ for the unit respondents already might be close to or even exceed $N$.

We avoid this undesirable outcome by down-weighting each $w_i^a$ to attempt to get close to $w_i^d$.  In some instances, enough information about the weighting adjustment procedures is provided to enable analysts to do so exactly. Often, however, this is not the case.
Absent information to allow reverse-engineering of the design weights, analysts can use an ad hoc and simple technique as follows. As before, set $w_i^a = 0$ whenever $U_i = 1$.  For all $i \in \mathcal{D}$, we create and use for estimation, 
\begin{equation}
w_i^* =  \left\{\begin{array}{lr}
        w_i^a\times (1-\frac{\sum_{j \in \mathcal{D}} U_j}{n}), & \text{if } U_i = 0\\
        \frac{\sum_{j \in \mathcal{D}} w_j^a}{n}, & \text{if } U_i = 1.
        \end{array}\right.
\label{approx_designW}
\end{equation}
Here, \eqref{approx_designW} down-weights the adjusted weights for unit respondents by the unit response rate, and assigns the remaining weight evenly to the unit nonrespondents. After this re-assignment step, $\sum_{i \in \mathcal{D}} w_i^*  = \sum_{i \in \mathcal{D}} w_i^a$, which should be a reasonable estimate of $N$. 

Analysts could create weights for unit nonrespondents by instead requiring $\sum w_i^* = N$. 
We find in simulation studies that setting $w_i^*$ so that $\sum w_i^* = N$ or $\sum w_i^* = \sum w_i^a$ does not make much difference to the ultimate inferences \citep{TangDis}. 

\subsection{Estimation Strategy for Illustrative Model}
\label{MCMC}
In this section, we illustrate a general estimation strategy for the  hybrid missingness MD-AM model using the model defined by \eqref{mix_Rx}---\eqref{mix_x1_u}.  The strategy can be extended to handle additional variables by replicating the various sampling steps.  We present the  sampler when design weights for unit respondents are published. When those weights are not available, we substitute $w_i^*$ for $w_i$.  We assume $T_{X_1}$ and  $T_{X_2}$ are available in $\mathcal{A}$, and that the analyst has specified ${V}_{X_1}$ and ${V}_{X_2}$.

We use normal approximations for the distributions of the estimated coefficients of the logistic regressions to simplify computations \citep{Raghunathan2001}. Drawing from the exact posterior distribution can be achieved via data augmentation using P\'{o}lya-Gamma latent variables \citep{Polson2013}, although with large enough samples the normal approximations often are reasonable.

Let $n_U$ be the number of unit nonrespondents in $\mathcal{D}$.  
 We get starting values for the sampler by imputing $X_{i1}$ for units with $R_{i1}^x=1$ under the assumption of MAR. We then compute maximum likelihood estimates for all parameters in the logistic regressions from this completed data set, which we use as starting values. The sampler proceeds as follows from any iteration $t$.  
  
\begin{enumerate}[label=IM\arabic*.]
\item Draw a value $\hat{T}_{X_2}^{(t+1)} \sim N(T_{X_2}, {V}_{X_2})$.
\item Calculate the number of times $X_{i2}^{(t+1)}$ should be imputed as 1 when $U_i = 1$ so that the weighted sum of $X_2^{(t+1)}$ is as close to $\hat{T}_{X_2}^{(t+1)}$ as possible; denote this number as $n_{2U}$. Specifically,
\begin{equation}
n_{2U} = \lfloor \frac{\hat{T}_{X_2}^{(t+1)} -\sum_{i \in \mathcal{D}} w_iX_{i2}^{(t+1)}I(U_i = 0)}{\sum_{i \in \mathcal{D}} w_i I(U_i = 1)/n_U} \rfloor.
\end{equation}
\item Let $\hat{\alpha}_0$ and $V_{\hat{\alpha}_0}$ denote the maximum likelihood estimate and the variance of $\alpha_0$ based on the $U_i = 0$ observations. Sample $\alpha_0^{(t+1)}$ from its approximate posterior distribution, $N(\hat{\alpha}_0,V_{\hat{\alpha}_0})$.
\item Calculate the proportion of unit nonrespondents that should be imputed as 1 for $X_2$. Denote this number as $\hat{p_2} = \frac{n_{2U}}{n_U}$. Then, set $\alpha_1^{(t+1)}$ so that $\mathbb{E}[\alpha_0^{(t+1)}+ \alpha_1^{(t+1)}] = logit(\hat{p}_2)$.
\item Draw imputations of $X_2$ for unit nonrespondents from $X_2|U=1 \sim Bernoulli\Big(1/\big(1 + \exp(-\alpha_0^{(t+1)} - \alpha_1^{(t+1)})\big)\Big)$.

\item Draw a value $\hat{T}_{X_1}^{(t+1)} \sim N(T_{X1}, {V}_{X_1})$. 
\item Calculate the number of times $X_{i1}^{(t+1)}$ should be imputed as 1 when $U_i = 1$ so that the weighted sum of $X_1^{(t+1)}$ is as close to $\hat{T}_{X1}^{(t+1)}$ as possible; denote this number as $n_{1U}$. Specifically,
\begin{equation}
n_{1U} = \lfloor \frac{\hat{T}_{X_1}^{(t+1)} - \sum_{i}^n w_iX_{i1}^{(t)}I(U_i = 0)}{\sum_{i}^n w_i I(U_i = 1)/n_U} \rfloor.
\end{equation}
\item Let $\hat{\beta}$ and $V_{\hat{\beta}}$ denote the maximum likelihood estimates and the covariance matrix of $\begin{bmatrix}
\beta_0 \\
\beta_1
\end{bmatrix}$ based on the $U_i = 0$ observations. Sample $\begin{bmatrix}
\beta_0 \\
\beta_1
\end{bmatrix}^{(t+1)}$ from its approximate posterior distribution $N(\hat{\beta},V_{\hat{\beta}})$.
\item Calculate the proportion of unit nonrespondents that should be imputed as 1 for $X_1$. Denote it as $\hat{p_1} = \frac{n_{1U}}{n_U}$. Then, set $\beta_2^{(t+1)}$ so that $\mathbb{E}[\beta_0^{(t+1)}+\beta_1^{(t+1)}X_2 + \beta_2^{(t+1)}] = logit(\hat{p}_1)$.
\item Draw imputations of $X_1$ for unit nonrespondents from $X_1|U=1 \sim Bernoulli\Big(1/\big(1+\exp(-\beta_0^{(t+1)}-\beta_1^{(t+1)}X_2 - \beta_2^{(t+1)})\big)\Big)$.
\item Let $\hat{\gamma}$ and $V_{\hat{\gamma}}$ denote the maximum likelihood estimates and the covariance matrix of $\begin{bmatrix}
\gamma_0 \\
\gamma_1 
\end{bmatrix}$ based on $U_i = 0$ observations. Sample $\begin{bmatrix}
\gamma_0 \\
\gamma_1 
\end{bmatrix}^{(t+1)}$ from its approximate posterior distribution, $N(\hat{\gamma},V_{\hat{\gamma}})$.\\
\item Draw imputations of $X_1^{(t+1)}$ when $(R_{i1}^x = 1, U_i = 0)$ from a $
Bernoulli\Big(1/\big(1+\exp(-\beta_0^{(t+1)}-\beta_1^{(t+1)}X_2\big)\Big)$.
\end{enumerate}
We update $X_2$ first and use it as a predictor of $X_1$ because, in our illustrative setting, $X_2$ only suffers from unit nonresponse whereas $X_1$ also suffers from item nonresponse.  When both $X_2$ and $X_1$ have item nonresponse, analysts can generate starting values for both using MICE. 
In general, we recommend updating variables with smaller item nonresponse rates earlier in the sequence.

This algorithm can be viewed as a generalized version of the one proposed in \cite{Pham2019}, which incorporates an offset called the ``calibrated-$\delta$ adjustment'' calculated from auxiliary information. The algorithm in \cite{Pham2019} handles item nonresponse only, whereas our algorithm, which we call the intercept matching algorithm, handles both item and unit nonresponse.


The intercept matching algorithm can be readily generalized to more than two variables with auxiliary margins, assuming we use the margins to add the indicator $U$ to the models for the survey variables. This can be done via repeating IM1--IM5 or IM6--IM10: generate parameter draws from their full conditional based on the $U_i = 0$ cases alone, use the auxiliary margin to adjust the intercept, and  draw imputations for this variable for unit nonrespondents using the adjusted intercept. When the model for $X_k$ is a multinomial logistic regression, as is the case for some variables in the 2018 CPS hybrid missingness MD-AM model, we estimate parameters and set $U$ coefficients in the logit expression for each level of $X_k$. 

The algorithm also easily incorporates variables without known margins. We add a conditional model for each  $Y_k$ to the collection, without including  terms for $U$. For example, for a binary $Y_k$, we follow IM3 (or IM8) to generate parameter draws based on the $U_i = 0$ cases alone.  We skip IM4 (or IM9) as there is no auxiliary margin associated with $Y_k$ for the intercept adjustment. As a result, we draw imputations of $Y_k$ for unit nonrespondents from the model used for unit respondents,  without changing the intercept. When $Y_k$ also suffers from item nonresponse, we follow IM11--IM12 to draw imputations for cases with $R_{ik}^Y=1$. In general, to facilitate modeling, we recommend specifying models first for $U$ and $X \mid U$, followed by models for $Y \mid X$, $R^x$, and $R^y$.  

\subsection{Summary of Simulation Results}
\label{simsummary}

In the supplementary material, we present results of a series of simulations that investigate the performance of the intercept matching algorithm for fitting the hybrid missingness MD-AM model with unequal probability samples subject to unit and item nonresponse. The simulations vary the departure from ignorability for the unit and item nonresponse, and the strength of association between $X_1$ and $X_2$. To conserve space, here we summarize the main findings from those simulations.  

We begin with the simulations that assume the design weights for unit respondents are known.  Across all scenarios we investigated, the  MD-AM model fit with the intercept matching algorithm  has better repeated sampling properties than default ICIN models that do not utilize the auxiliary margins.  In particular, when missingness is nonigorable, the MD-AM models generally result in multiple imputation point estimates with much lower bias than the default ICIN models.  Confidence interval coverage rates for the MD-AM models generally are close to or exceed 95\%, whereas those for the default ICIN model often have zero coverage due to the bias. Interestingly, the multiple imputation variance estimates with the MD-AM model are positively biased.  This mirrors a finding in \cite{Jerry2008}, who noted that the multiple imputation variance estimator of \citet{Rubin1987} has positive bias when the imputation models use more information---in this case, the auxiliary margins---than the analysis models. 

When the design weights for unit respondents are not known, so that we use \eqref{approx_designW} to make analysis weights, the performance of the MD-AM model suffers compared to when we  can use \eqref{weight_UN}.  This is because we are forced to  estimate the design weights, and errors in those estimates carry through to the survey analysis.  Nonetheless, the point and interval estimates from the MD-AM model continue to outperform those from the default ICIN model. In fact, for settings where the bias due to nonresponse is not extreme, the MD-AM model produces results with good repeated sampling properties.  We refer the reader to the supplementary material for details.

The bottom line for the simulation studies is that the inferences from the hybrid missingness MD-AM  model with the intercept matching algorithm outperform those from default models that do not use the margins, even in the case where only adjusted weights are available.   These promising results encourage us to estimate a hybrid missingness MD-AM model with the intercept matching algorithm to analyze voter turnout in the 2018 CPS data, to which we now turn.

\section{Estimating Voter Turnout with 2018 CPS Data}
\label{CPS}
In every Congressional and Presidential election year since 1964,  the Voting and Registration Supplement has been included biennially in the November basic monthly survey of the Current Population Survey. 
The resulting data are regarded as one of the premier sources for studying turnout in the U.\ S.  The large sample size and detailed demographic information enable researchers to examine voter turnout by state and 
within subgroups, for example, among racial and ethnic minority groups \citep{CPS_bias}. 

In spite of its status as a gold standard for research on turnout, the CPS suffers from unit and item nonresponse. To handle unit nonresponse, the Census Bureau adjusts the design weights of CPS survey respondents so that the resulting weighted estimates match certain demographic totals. To handle item nonresponse, the Census Bureau uses various imputation methods.  Particularly relevant for turnout studies, the Census Bureau imputes all ``Don't Know,'' ``Refused,'' and ``No Response'' answers to the question on turnout as non-voters.  Taken together, the Census Bureau's treatment of missingness represents very strong modeling assumptions \citep{Hur2013}.  To be explicit, according to the Census Bureau missingness assumptions, unit nonresponse is conditionally independent of voting status, and item nonresponse on vote perfectly predicts voting status.  

We instead handle the unit and item nonresponse with a hybrid missingness MD-AM model.  We note that \citet{Akande2021MDAM} previously analyzed CPS turnout data with an MD-AM model.  Their analysis uses data from the 2012 CPS, treats the CPS as a simple random sample (i.e., disregards the survey weights), and uses selection models for all nonresponse indicators rather than the hybrid missingness mechanism.


\subsection{Data}
We use 2018 CPS data for North Carolina downloaded from the IPUMS website (\url{https://cps.ipums.org/cps/}) and the variables described in Table \ref{variable:description}.
The data are collected via  a multi-stage design, including stratification within states, probability proportional to size sampling, and oversampling of certain demographic populations.  As a result, the design weights vary across sampled units.  This variation  can be compounded by the weighting adjustments for unit nonresponse.

The CPS reports unit nonresponse rates at the household level, that is, whether or not the household responds, rather than the individual person level.  We therefore need to approximate the person level unit nonresponse rate among those eligible for the November supplement  (U.\ S.\ citizens at least 18 years old).  To do so, we use the strategy developed by \citet{Akande2021MDAM}, which we summarize here. First, using information from the CPS data file describing why sampled households failed to respond, we exclude ``Type C'' households; these are deemed ineligible for the survey. 
Second, we estimate the average number of adult citizens per household in North Carolina. Our numerator derives from the Census Bureau's special tabulation of the Citizen Voting Age Population (CVAP) using five-year American Community Survey (ACS) data in North Carolina, and our denominator is the estimated total number of households in North Carolina also from the five-year ACS data. Third, we multiply the number of eligible nonresponding households in the CPS data by the estimated average number of adult citizens per household, and round the product to the nearest person. We then append records to the observed CPS data comprising no information (other than that they reside in North Carolina and are eligible to vote).

\begin{table}[t]
	\caption{\label{variable:description}Description of variables used in the analysis of turnout in the 2018 CPS data, along with notation used for modeling.}
	\footnotesize
	\centering
	\begin{tabular}[c]{lcl}
		\toprule
		Variable  & Notation  & Categories \\
		\midrule
		Sex & S & 0 = Male, 1 = Female \\
Race & E &  1 = White alone, 2 = Black alone, 3 = Hispanic or Latino, 4 = Remaining people\\ 
		Education & C & 1 = High school or less (HS-), 2 = Some college, 3 = Bachelor's and more (BS+)\\
		Age & A & 1 = 18 - 29, 2 = 30 - 39, 3 = 40 - 49, 4 = 50 - 59, 5 = 60 - 69, 6 = 70 - 79, 7 = 80+ \\
		Vote & V & 0 = Did not vote; 1 = Voted \\
		\bottomrule
	\end{tabular}
\end{table}

Using this approach, we generate 913 unit nonrespondents to add to the 2013 unit respondents; thus, we have $n=2926$ in $\mathcal{D}$ with  31\% unit nonresponse. Two observations in the CPS data have all variables missing except their survey weights. We treat them as unit nonrespondents, without modifying their weights. Item nonresponse rates among respondents are modest among sex (0\%), race (2\%), education (4\%), and age (4\%).  Item nonresponse among respondents for vote is more substantial at 18\%.

We use auxiliary margins for several of the variables in Table \ref{variable:description}.  For vote, we 
 use the total ballots counted voter-eligible population (VEP)  turnout rate for North Carolina in the 2018 election, which is 49\% (\url{http://www.electproject.org/2018g}).  For sex and race, we utilize auxiliary margins for the VEP from the 2018 ACS for North Carolina.  Specifically, we have 52\% female and 48\% male; we have 69.9\% with white race alone, 21.8\% with black race alone, 3.9\% with Hispanic, and 4.4\% comprising people not in these categories. The standard errors of these statewide ACS estimates are small enough that we can treat them as known population margins.  


\subsection{Hybrid missingness MD-AM model}
\label{CPS_model}
For any individual $i \in \mathcal{D}$, let $A_i$ denote their age, $S_i$ denote their sex, $E_i$ denote their race/ethnicity, $C_i$ denote their education, and $V_i$ denote their vote.  We use this notation instead of $X_{ik}$ and $Y_{ik}$ to make it easier to identify the variables. Note that $\{G,E,V\} \in \mathcal{A}$ and $\{A,C\} \notin \mathcal{A}$. 
For any individual $i \in \mathcal{D}$, let $(U_i, R^E_i, R^C_i,  R^A_i, R^V_i)$ represent the indicators for unit nonresponse and item nonresponse for race/ethnicity, education, age, and vote, respectively.  

Following the MD-AM framework, we first specify an identifiable model for the collection of survey variables and nonresponse indicators without using auxiliary margins.   Table \ref{tab:model} summarizes the sequence of conditional models before using the margins. 
Exploratory analysis of the missing data patterns reveals that whenever an individual's age is missing, that person's vote also is missing. Hence, when specifying the default model for $R^A$, we cannot include a term for $V$; there is not sufficient information to identify a coefficient of vote. We set $R^V_i= 1$ whenever $R^A_i= 1$.  In the model for $V$, we include  interactions between $S$ and $E$, $S$ and $C$, and $S$ and $A$ as we are especially interested in estimating voter turnout in these subgroups.  Overall, the collection of models specifies fewer parameters than can be identified using the CPS data alone; however, the simplified models explain the data reasonably well while avoiding the need to estimate extra parameters, especially in the multinomial models. 

\begin{table}[t]
	\caption{\label{tab:model}Hybrid missingness MD-AM model in the analysis of turnout in the 2018 CPS data.  Abbreviations for models include ``B'' for Bernoulli distribution, ``MR'' for multinomial logistic regression, ``LR'' for logistic regression.  For the predictors, ``I'' stands for an intercept. Categorical predictors with $d$ levels are modeled with a series of $d-1$ indicator variables. Interaction effects between any two variables $A$ and $B$ represented by $A:B.$}
	\footnotesize
	\centering
	\begin{tabular}[c]{lcl}
		\toprule
		Variable  & Model & Predictors \\
		\midrule
$U$ & B & $I$ \\
		$S$ & B & $I$ \\
 $E$ &  MR &  $I + S$\\ 
		$C$ & MR & $I + S + E$\\
		$A$ & MR & $I + S + E + C$ \\
		$V$ & LR & $I + S + E + C + A + S:E + S:C + S:A$ \\
	$R^E$ & LR & $I + S + C + A + V$ \\
		$R^C$ & LR & $I + S + E + A + V$ \\
		$R^A$ & LR & $I + S + E + C$ \\
		$R^V$ & LR & $(R^A = 0):  I + S + E + C + A$ \\
		& & $(R^A = 1): R^V=1$\\
		\bottomrule
	\end{tabular}
\end{table}

Following step 2 of the MD-AM framework with hybrid missingness, we add main effects for $U$ to the models for $S$, $E$, and $V$. We do not add $U$ to the models for $A$ and $C$ since we do not utilize auxiliary margins on age and education.

As the CPS file from IPUMS includes only adjusted weights, we apply the algorithm in \eqref{approx_designW} to create the analysis weights. We use non-informative priors for all model parameters and normal approximations for the coefficients of all regressions.  We employ  the intercept matching algorithm 
for 10000 iterations and discard the first 5000 iterations as burn-in. Using a non-optimized code in $R$ and a standard laptop, it takes about 14 hours to  generate the 10000 draws. Evaluations of trace plots of model parameters suggest that the sampler converges. Among the 5000 posterior samples, we retrieve every 100th posterior sample to create $L = 50$ multiple imputation data sets, and do survey-weighted multiple imputation inference using those completed data sets. 

\subsection{Results}
We begin by examining the completed-data marginal distributions of the variables other than vote, i.e.,  $S$, $E$, $C$ and $A$.  
Table \ref{marginal_CPS_Int} displays the multiple imputation point estimates, $\bar{q}_L$, and corresponding multiple imputation standard errors using the 50 completed data sets generated under the hybrid missingness MD-AM model.  Survey-weighted estimates are computed with the ``survey'' package \citep{surveyR} in $R$, assuming each completed data set is  a probability proportional to size with replacement sample. This is a common estimation strategy for analyzing complex surveys \citep{Lohr2010}.  For comparisons, we also generate 50 completed  data sets using a default implementation of MICE with only the unit respondents.  Estimates for the MICE-completed data sets are based on the CPS adjusted weights on the IPUMS file.  Finally, we also display weighted estimates based on complete cases.

\begin{table}[t]
\centering
\caption[Results from the CPS data analysis for marginal distributions.]{Estimated marginal distributions of sex, age, and race based on 50 imputations generated from the hybrid missingness MD-AM model and default MICE. The entries under the column labeled $Q$ are the available population percentages used as auxiliary margins. The entries in the columns labeled CC are survey-weighted estimates using the complete cases. 
}
\begin{tabular}{lrrrrrrr}
\hline
&  & \multicolumn{2}{c}{MD-AM} & \multicolumn{2}{c}{MICE} & \multicolumn{2}{c}{CC}\\
& $Q$ & $\bar{q}_L$ & SE  & $\bar{q}_L$ & SE &  Est & SE\\
\hline
Male & .48 & .478 & .017  & .470 & .011 & .466 & .013\\
Female & .52 & .522 & .017  & .530 & .011 & .534 & .013\\
\hline
White & .699 &  .699 & .013 & .696 & .011 & .709 & .012\\
Black & .218 &  .219 & .013  & .220 & .010 & .213 & .011\\
Hispanic & .039 & .038 & .005  & .035 & .004 & .033 & .005\\
Rest & .044 & .044 & .005  & .049 & .005 & .046 & .005\\
\hline
$(0,29]$ & & .213 & .014 & .216 & .010 & .205 & .011\\
$(29,39]$ & & .160 & .021  & .141 & .008 & .144 & .009\\
$(39,49]$ & & .189 & .018  & .161 & .009 & .150 & .009\\
$(49,59]$ & & .162 & .013  & .173 & .009 & .177 & .009\\
$(59,69]$ & & .147 & .010  & .161 & .008 & .168 & .009\\
$(69,79]$ & & .092 & .009  & .103 & .007 & .111 & .008\\
$>79$ & & .036 & .005 &  .043 & .004 & .045 & .005\\
\hline
HS- & & .372 & .012 &  .375 & .011 & .366 & .012\\
Some College & & .297 & .011  & .300 & .011 & .306 & .012 \\
BA+ & & .331 & .012 &  .325 & .011 & .328 & .012\\
\hline
\end{tabular}
\label{marginal_CPS_Int}
\end{table}

\begin{table}[h!]
\centering
\caption[Proportion who voted in each sub-group of the PPS data.]{Estimated proportion who voted in subgroups based on 50 imputations generated from the hybrid missingness MD-AM model and default MICE. Survey-weighted estimates for complete case analyses (CC) and estimates released by the Census Bureau (Census) also included.   The auxiliary margin for voted in North Carolina is .49.}
\begin{tabular}{lrrrrrrrrr}
\hline
& \multicolumn{2}{c}{MD-AM} & \multicolumn{2}{c}{MICE} & \multicolumn{2}{c}{CC} & \multicolumn{2}{c}{Census}\\
\hline
 & $\bar{q}_L$ & SD & $\bar{q}_L$ & SD & Est & SD & Est & SD\\
\hline
Full & .501 & .016  & .630 & .012 & .637 & .012 & .524 & .012\\
Male & .499 & .021  & .630 & .018 & .639 & .018 & .521 & .017\\
Female & .504 & .020 & .631 & .017 & .636 & .017 & .526 & .016\\
\hline
White &  .510 & .018 & .645 & .014 & .648 & .014 & .544 & .013\\
Black &  .519 & .032  & .647 & .031 & .658 & .030 & .516 & .027 \\
Hispanic & .390 & .061  & .494 & .070 & .489 & .073 & .386 & .061\\
Rest & .378 & .057  & .449 & .058 & .467& .060 & .371 & .049 \\
\hline
$(0,29]$ & .325 & .027  & .419 & .030 & .421 & .030 & .331 & .026\\
$(29,39]$ & .398 & .038  & .558 & .033 & .571 & .034 & .466 & .031\\
$(39,49]$ & .492 & .041  & .679 & .029 & .681 & .030 & .517 & .028\\
$(49,59]$ & .595 & .038  & .698 & .027 & .700 & .027 & .596 & .026 \\
$(59,69]$ & .700 & .035  & .796 & .024 & .794 & .024 & .664 & .025\\
$(69,79]$ & .672 & .039  & .758 & .030 & .761 & .030 & .691 & .031\\
$>79$ & .467 & .060  & .545 & .055 & .556 & .054 & .485 & .051\\
\hline
HS- &  .347 & .020 & .456 & .021 & .464 & .021 & .369 & .018\\
Some College & .519 & .023  & .653 & .022 & .653 & .022 & .552 & .021 \\
BA+ & .659 & .029  & .810 & .017 & .813 & .017 & .676 & .018\\
\hline
\end{tabular}
\label{one_sub_Int}
\end{table}

\begin{table}[h!]
\centering
\caption[Proportion who voted in each subgroups of gender crossed with race/age of the CPS data.]{Estimated proportion who voted in detailed subgroups based on 50 imputations generated from the  hybrid missingness MD-AM model  and default MICE.   Survey-weighted estimates for complete case analyses (CC) and estimates released by the Census Bureau (Census) also included.  The auxiliary margin for voted in North Carolina is .49.}
\begin{tabular}{lrrrrrrrrr}
\hline
& \multicolumn{2}{c}{MD-AM} & \multicolumn{2}{c}{MICE} & \multicolumn{2}{c}{CC} & \multicolumn{2}{c}{Census}\\
\hline
 & $\bar{q}_L$ & SD & $\bar{q}_L$ & SD & Est & SD & Est & SD \\
\hline
Male, White &  .513 & .023 &  .648 & .020 & .650 & .020 & .550 & .040\\
Male, Black &  .475 & .047  & .612 & .046 & .618 & .045 & .463 & .018\\
Male, Hispanic & .459 & .100  & .562 & .103 & .597 & .106 & .443 & .086\\
Male, Rest & .427 & .081  & .509 & .086 & .552 & .090 & .416 & .074\\
Female, White &  .508 & .023 & .642 & .019 & .647 & .019 & .539 & .018\\
Female, Black &  .557 & .038  & .676 & .039 & .687 & .040 & .559 & .037\\
Female, Hispanic & .323 & .080  & .432 & .094 & .401 & .100 & .329 & .085\\
Female, Rest & .334 & .072  & .396 & .076 & .397 & .078 & .332 & .066\\
\hline
Male, $(0,29]$ & .313 & .038  & .405 & .042 & .408 & .043 & .317 & .036 \\
Male, $(29,39]$ & .396 & .049  & .553 & .048 & .573 & .049 & .460 & .045 \\
Male, $(39,49]$ & .451 & .053  & .640 & .045 & .639 & .046 & .482 & .042\\
Male, $(49,59]$ & .577 & .048  & .694 & .037 & .689 & .038 & .596 & .037\\
Male, $(59,69]$ & .714 & .049  & .813 & .034 & .810 & .035 & .656 & .037\\
Male, $(69,79]$ & .764 & .050  & .828 & .040 & .840 & .039 & .781 & .042\\
Male, $>79$ & .460 & .094  & .561 & .090 & .565 & .091 & .469 & .083 \\
Female, $(0,29]$ & .337 & .038  & .433 & .041 & .432 & .043 & .344 & .037 \\
Female, $(29,39]$ & .401 & .050  & .562 & .045 & .569 & .046 & .472 & .042\\
Female, $(39,49]$ & .530 & .048  & .710 & .037 & .714 & .038 & .545 & .037\\
Female, $(49,59]$ & .594 & .046  & .702 & .039 & .711 & .037 & .596 & .037\\
Female, $(59,69]$ & .687 & .043  & .780 & .034 & .781 & .034 & .671 & .035 \\
Female, $(69,79]$ & .596 & .054  & .700 & .044 & .694 & .044 & .619 & .044\\
Female, $>79$ & .473 & .072  & .535 & .068 & .551 & .068 & .494 & .065\\
\hline
Male, HS- & .347 & .026 &  .459 & .029 & .468 & .030 & .369 & .025\\
Male, Some College & .520 & .035  & .660 & .032 & .656 & .033 & .552 & .031\\
Male, BA+ & .677 & .039  & .835 & .025 & .837 & .024 & .699 & .027\\
Female, HS- &  .348 & .026  & .452 & .030 & .460 & .029 & .370 & .026\\
Female, Some College & .519 & .031  & .648 & .030 & .650 & .030 & .552 & .029\\
Female, BA+ & .645 & .034  & .791 & .024 & .794 & .024 & .658 & .025\\
\hline
White, HS- & .341 & .023 & .445 & .022 & .452 & .025 & .374 & 022 \\
White, Some College & .512 & .026 & .647 & .024 & .654 & .026 & .556 & .024 \\
White, BA+ & .663 & .030 & .812 & .017 & .826 & .018 & .694 & .019 \\
White, Male, HS- & .341 & .031 & .452 & .035 & .453 & .035 & .378 & .031 \\ 
\hline
\end{tabular}
\label{two_sub_Int}
\end{table}

Overall, estimates are similar across all the survey-weighted analyses. This is not surprising, as the CPS adjusts design weights so that the survey-weighted estimates for some of these demographic variables approximately match their known demographic margins. 
We point out that,  for $S$ and $E$ where we have auxiliary margins, the multiple imputation standard deviations for the MD-AM model are biased high. The true standard deviation for each of these two $\bar{q}_L$ is close to zero, as the model essentially forces each of these two $\bar{q}_L$ to match their corresponding known margins.  

The benefits of handling item and unit nonresponse with the hybrid missingness MD-AM model become apparent when we examine voter turnout.  Table \ref{one_sub_Int} displays estimates of turnout by each category of the demographic variables for the three analysis strategies.  Additionally,  Table \ref{one_sub_Int} includes the turnout estimates using the Census Bureau procedures for handling missing values.
That is, we use the CPS data for unit respondents and their survey weights on the file; we use the Census Bureau's imputations for $S$, $E$, $A$, and $C$; and, we set all missing responses for $V$ as not voting. 

Statewide, the MD-AM model estimates around 50.1\% of CPS participants are voters, which is close to the population turnout of 49\%.  In contrast, estimates from MICE and complete cases overestimate the turnout rate overall  and seemingly for most subgroups---a bias likely more than 10 percentage points in many instances.  For subgroups, estimates of turnout under the MD-AM model tend to be more plausible than those under MICE and the complete case analyses.  For example, college graduates are estimated to have a turnout rate of more than 80\% using complete case or MICE analysis, where as the MD-AM estimates a  turnout rate around 66\%. 

Table \ref{two_sub_Int} displays turnout for bivariate combinations of the demographic variables. Here again, the MD-AM model estimates are smaller and more plausible than those under MICE and complete case analysis.  For some of the most substantively interesting population subgroups, the MD-AM estimates very different turnout rates than the MICE or complete case analyses.  For example, political pundits have focused considerable attention on voting behavior of non-college educated white men and women.  Whereas the complete case analysis estimated they had a turnout rate of 45\% in 2018, the MD-AM estimates their turnout rate at just 34\%. 
As another example, the MD-AM model estimates that Black men had a much lower turnout rate (47.5\%, which is in line with the statewide turnout rate) compared to the complete case estimates (61.8\%, over 10 points higher than the statewide rate).


The results from the MD-AM model and Census Bureau imputations are somewhat similar.  By imputing all item nonrespondents on vote to be non-voters, the Census Bureau essentially is forcing the overall turnout margin to be closer to 49\% than the complete case results. Still, the overall turnout estimates under the MD-AM model are closer to 49\%.  Additionally, it seems implausible that all item nonrespondents are non-voters. In contrast, the MD-AM model imputes on average 60\% of the item nonrespondents as voters---close to the complete cases turnout estimate of 63.7\%, as expected from the ICIN assumption for $R^V$---and, to make the completed-data design-based estimates of turnout closely match 49\%, on average 22\% of the unit nonrespondents as voters. 

\begin{figure}[t]
\centering
\includegraphics[scale=0.5]{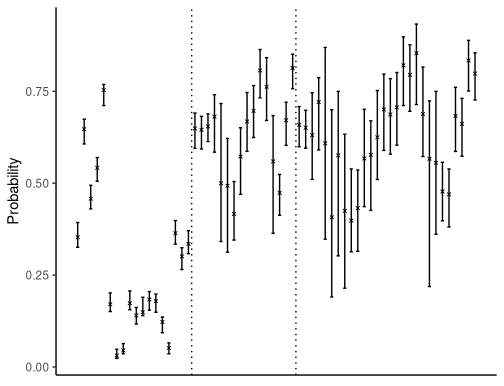}
\caption[Posterior predictive intervals for all marginal and conditional probabilities involving vote for completed data only.]{Posterior predictive intervals for all marginal and conditional probabilities involving vote for completed data only. Crosses are observed data estimates. The first panel shows marginal probabilities of vote, sex, race/ethnicity, age and education. The second panel shows voter turnout probabilities in various sex, race/ethnicity, age and education subgroups. The third panel displays voter turnout probabilities in sex crossed with race/ethnicity, gender crossed with age and sex crossed with education subgroups.}
\label{post_check_Int}
\end{figure}

While it is impossible to assess how accurately the MD-AM model describes the nonresponse mechanism in the CPS, we can assess how well the MD-AM model reproduces the observed data. To do so, we compare unweighted, observed percentages from $\mathcal{D}$ with estimates from replicated data generated from posterior predictive distributions \citep{He_Alan_2012}.  These comprise the values of $\{(A_i, S_i, E_i,C_i, V_i): U_i= R^A_i =  R_i^E= R_i^C = R_i^V=0\}$. Using each of the 5000 draws of the parameters, we generate new values of all variables for all $n$ units in $\mathcal{D}$, and compute the replicated data percentages for those generated to be respondents to all questions. We construct 95\% intervals  for marginal probabilities for the replicated data 
for all variables and voter turnout probabilities for subgroups. 
Figure \ref{post_check_Int} displays the posterior predictive intervals.
All intervals contain the observed data estimates, giving us confidence in the reasonableness of the hybrid missingness MD-AM model for the observable portion of the data. 

The simulation studies in the supplementary material suggest that, since design weights for unit respondents are not available, the performance of the MD-AM model in estimating voter turnout should depend on the strength of the unit nonresponse bias. 
The similarity of estimates in Table \ref{marginal_CPS_Int} for the MD-AM model, MICE and complete case analysis suggest practically irrelevant biases for these three variables  due to unit nonresponse (after the weighting adjustments). On the contrary, the voter turnout estimated under the MD-AM model is lower than the turnout under MICE and complete case analysis, suggesting the unit nonresponse bias on vote is important.  The simulation results under such scenarios in the supplementary material  (namely, scenarios ``b" or ``d" in Table 8 in the supplement) suggest that  results from the MD-AM model are more accurate than those from models that do not  utilize the information in the margins. 

\subsection{Incorporating measurement error}\label{CPSmeasurementerror}
Beyond the effects of item and unit nonresponse, it is well-recognized that self-reported voter turnout suffers from measurement error. More specifically, people often report that they voted when they did not in fact do so.  In contrast, people who did vote rarely report that they did not \citep[]{Enamorado2019, jackman2019does, debell2020turnout}.  Thus, some of the apparent bias in the CPS complete cases results  could be the result of over-reporting of turnout.

\begin{figure}[t]
    \centering
    \includegraphics[scale=0.5]{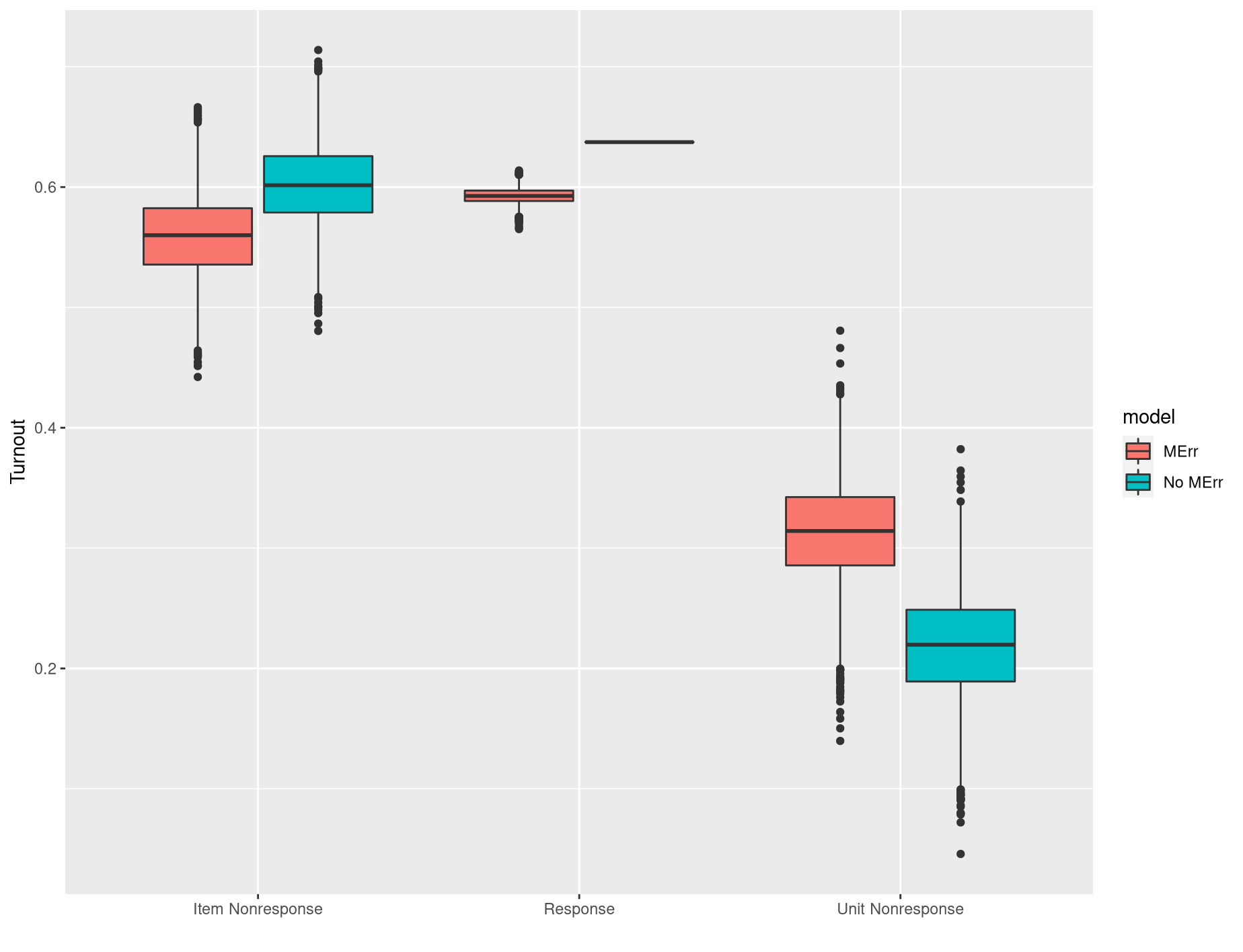}
    \caption{Posterior distribution of voter turnout among item nonrespondents, respondents, and unit nonrespondents for the MD-AM model.  For each response type, the box to the right represents the turnout rate in the MD-AM model assuming no measurement error among respondents' vote answers.  The box to the left represents the turnout rate in the MD-AM model assuming measurement error.}
    \label{MErr_fig}
\end{figure}

To assess the sensitivity of results to reporting error, we layer a measurement error model on top of the  hybrid missingness MD-AM  model.  Previous research  suggests a dose-response relationship between over-reporting of vote and  education: increasing education is associated with increasing rates of over-reporting \citep{Silver_Anderson_Abramson_1986,Karp_Brockington_2005}. 
Unfortunately, we do not have a validated sample from the CPS to assess the over-reporting rate by educational attainment.  Thus, we represent the measurement error in vote using informative prior distributions.

For each individual $i \in \mathcal{D}$ among those who have a response to the question about vote, let $Z_i=1$ when the person is reported as a voter, and let $Z_i=0$ when the person is reported as a nonvoter.  We seek to tie each $Z_i$ to the underlying true voting status, $V_i$.
Note that we define $Z_i$ only  for units with $U_i = R^V_i =0$. Let the misreporting probabilities for nonvoters given their education level be as follows.
\begin{align}
    \mathbb{P}(Z_i=1|V_i=0,C_i = 1, U_i=R^V_i=0) & = \theta_1 \\
    \mathbb{P}(Z_i=1|V_i=0,C_i = 2, U_i=R^V_i=0) & = \theta_2 \\
    \mathbb{P}(Z_i=1|V_i=0,C_i = 3, U_i=R^V_i=0) & = \theta_3,
\end{align}
where $C_i \in \{1,2,3\}$ per Table \ref{variable:description}. 
 Among nonvoters who have at most a  high school education, we assume around 6\% would misreport that they actually voted ($\theta_1$).  Among nonvoters with some college education, we assume around 13\% would misreport that they actually voted ($\theta_2$). Among nonvoters who have at least a bachelor's degree, we assume around 19\% would misreport that they voted ($\theta_3$).  We treat these best guesses as measures of central tendency in Beta prior distributions.  
For each $\theta_c$, we assume $\theta_c \sim Beta(a_c,b_c)$, 
where $(a_1, b_1) = (60, 940)$, $(a_2, b_2) = (130, 870)$, and  $(a_3, b_3) = (190, 810)$.  These correspond to prior means of misreporting rates of $(0.06, 0.13, 0.19)$, respectively, and prior standard deviations around 1\%. In accordance with past research from other studies, we assume that anyone who truly voted is reported that they did so. We select these prior distributions based on the research findings from previous voter validation studies that have matched survey respondents to voter files \citep[e.g., ][]{jackman2019does}, accounting for differences in observed bias across election and survey type.  

Adding the measurement error models and prior distributions to the MD-AM model in Section \ref{CPS_model} requires  two additional steps in the intercept matching algorithm: update $\theta_c$, and sample true voting status $V_i$ for individuals with $(U_i = 0,  R^V_i=0, Z_i=1)$. The full conditionals for these updates are described in the supplementary material.

Figure \ref{MErr_fig} displays estimated voter turnout rates among respondents,  item nonrespondents on vote, and unit nonrespondents before and after taking the measurement error into consideration.  With the additional measurement error assumptions, the MD-AM model changes some of the reported voters to nonvoters, lowering the voting proportion among observed cases. Under the ICIN model for $R^V$, we therefore  impute fewer voters among  respondents with $R^V_i = 1$. Consequently, we need to impute more voters among unit nonrespondents to match the auxiliary margin for vote. 

Tables containing results from the hybrid missingness MD-AM model with measurement error like those in Table \ref{marginal_CPS_Int}, Table \ref{one_sub_Int}, and Table \ref{two_sub_Int} are displayed in the supplementary material. Overall, despite the measurement error, the estimates are quite similar with a few estimates increasing or decreasing by a point or two, and all changes are well within one estimated standard error.   


\section{Concluding Remarks}\label{conclusion}
The  hybrid missingness MD-AM model allows analysts to take advantage of known auxiliary margins when imputing values for unit and item  nonresponse in surveys with complex sampling designs.  In the 2018 CPS analysis, we enriched the models to account for unit nonresponse, as it was more prevalent than item nonresponse.  In other situations, analysts may prefer to enrich one or more of the models for the item nonresponse indicators.  We explored adding an indicator for $V$ to the model for $R^V$ and removing $U$ from the model for $V$ using a rejection sampler akin to that used in \cite{AkandeWeights}. However, with diverse values of weights, our sampler had a difficult time exploring the parameter space adequately.  Further research is needed on computational algorithms for the version of the hybrid missingness MD-AM model that enriches item nonresponse models.


The simulation studies demonstrate that the hybrid missingness MD-AM results are more accurate when we use the unit respondents' design weights as opposed to having to down-weight adjusted weights.  This finding suggests that survey providers would help analysts by including design weights in survey data files, even when they also include adjusted weights on those files.

\bibliographystyle{natbib}
\bibliography{lit}

\begin{thebibliography}{}

\bibitem[Akande and Reiter(2022)Akande and Reiter]{AkandeWeights}
Akande, O. and Reiter, J.~P. (2022).
\newblock Multiple imputation for nonignorable item nonresponse in complex
  surveys using auxiliary margins.
\newblock In A.~L. Carriquiry, J.~M. Tanur, and W.~F. Eddy, editors, {\em
  Statistics in the Public Interest: In Memory of Stephen E. Fienberg\/}, pages
  289--306. Springer, Berlin.

\bibitem[Akande {\em et~al.}(2021)Akande, Madson, Hillygus, and
  Reiter]{Akande2021MDAM}
Akande, O., Madson, G., Hillygus, D.~S., and Reiter, J.~P. (2021).
\newblock Leveraging auxiliary information on marginal distributions in
  nonignorable models for item and unit nonresponse in surveys.
\newblock {\em Journal of the Royal Statistical Society: Series A\/}, {\bf
  184}, 643--662.

\bibitem[Ansolabehere {\em et~al.}(2022)Ansolabehere, Fraga, and
  Schaffner]{CPS_bias}
Ansolabehere, S., Fraga, B.~L., and Schaffner, B.~F. (2022).
\newblock The {CPS} voting and registration supplement overstates minority
  turnout.
\newblock {\em Journal of Politics\/}, page forthcoming.

\bibitem[Azur {\em et~al.}(2011)Azur, Stuart, Frangakis, and Leaf]{MICE2011}
Azur, M.~J., Stuart, E.~A., Frangakis, C., and Leaf, P.~J. (2011).
\newblock Multiple imputation by chained equations: what is it and how does it
  work?
\newblock {\em International Journal of Methods in Psychiatric Research\/},
  {\bf 20}, 40--49.

\bibitem[DeBell {\em et~al.}(2020)DeBell, Krosnick, Gera, Yeager, and
  McDonald]{debell2020turnout}
DeBell, M., Krosnick, J.~A., Gera, K., Yeager, D.~S., and McDonald, M.~P.
  (2020).
\newblock The turnout gap in surveys: Explanations and solutions.
\newblock {\em Sociological Methods \& Research\/}, {\bf 49}(4), 1133--1162.

\bibitem[Deng {\em et~al.}(2013)Deng, Hillygus, Reiter, Si, and
  Zheng]{Deng2013}
Deng, Y., Hillygus, D.~S., Reiter, J.~P., Si, Y., and Zheng, S. (2013).
\newblock Handling attrition in longitudinal studies: the case for refreshment
  samples.
\newblock {\em Statistical Science\/}, {\bf 28}, 238--256.

\bibitem[Enamorado and Imai(2019)Enamorado and Imai]{Enamorado2019}
Enamorado, T. and Imai, K. (2019).
\newblock Validating self-reported turnout by linking public opinion surveys
  with administrative records.
\newblock {\em Public Opinion Quarterly\/}, {\bf 83}, 723--748.

\bibitem[Fuller(2009)Fuller]{Fuller_2009}
Fuller, W.~A. (2009).
\newblock {\em Sampling Statistics\/}.
\newblock John Wiley \& Sons, Inc, New York.

\bibitem[He and Zaslavsky(2012)He and Zaslavsky]{He_Alan_2012}
He, Y. and Zaslavsky, A.~M. (2012).
\newblock Diagnosing imputation models by applying target analyses to posterior
  replicates of completed data.
\newblock {\em Statistics in Medicine\/}, {\bf 31}, 1--18.

\bibitem[Hirano {\em et~al.}(2001)Hirano, Imbens, Ridder, and
  Rubin]{Hirano2001}
Hirano, K., Imbens, G.~W., Ridder, G., and Rubin, D.~B. (2001).
\newblock Combining panel data sets with attrition and refreshment samples.
\newblock {\em Econometrica\/}, {\bf 69}, 1645--1659.

\bibitem[Horvitz and Thompson(1952)Horvitz and Thompson]{HT1952}
Horvitz, D.~G. and Thompson, D.~J. (1952).
\newblock A generalization of sampling without replacement from a finite
  universe.
\newblock {\em Journal of the American Statistical Association\/}, {\bf 47},
  663--685.

\bibitem[Hur and Achen(2013)Hur and Achen]{Hur2013}
Hur, A. and Achen, C.~H. (2013).
\newblock Coding voter turnout responses in the current population survey.
\newblock {\em Public Opinion Quarterly\/}, {\bf 77}, 985--993.

\bibitem[Jackman and Spahn(2019)Jackman and Spahn]{jackman2019does}
Jackman, S. and Spahn, B. (2019).
\newblock Why does the american national election study overestimate voter
  turnout?
\newblock {\em Political Analysis\/}, {\bf 27}(2), 193--207.

\bibitem[Karp and Brockington(2005)Karp and Brockington]{Karp_Brockington_2005}
Karp, J.~A. and Brockington, D. (2005).
\newblock Social desirability and response validity: A comparative analysis of
  overreporting voter turnout in five countries.
\newblock {\em Journal of Politics\/}, {\bf 67}, 825--840.

\bibitem[Linero and Daniels(2018)Linero and Daniels]{linero:daniels}
Linero, A.~R. and Daniels, M.~J. (2018).
\newblock Bayesian approaches for missing not at random outcome data: The role
  of identifying restrictions.
\newblock {\em Statistical Science\/}, {\bf 33}, 198--213.

\bibitem[Lohr(2010)Lohr]{Lohr2010}
Lohr, S.~L. (2010).
\newblock {\em Sampling: Design and Analysis. 2nd Edition\/}.
\newblock Brooks/Cole, Boston.

\bibitem[Lumley(2021)Lumley]{surveyR}
Lumley, T. (2021).
\newblock {\em Analysis of complex survey samples\/}.
\newblock R ($>=$ 3.5.0), grid, methods, Matrix, survival.

\bibitem[Mealli and Rubin(2015)Mealli and Rubin]{Mealli_Rubin_2015}
Mealli, F. and Rubin, D.~B. (2015).
\newblock Clarifying missing at random and related definitions, and
  implications when coupled with exchangeability.
\newblock {\em Biometrika\/}, {\bf 102}, 995--1000.

\bibitem[Pham {\em et~al.}(2019)Pham, Carpenter, Morris, Wood, and
  Petersen]{Pham2019}
Pham, T.~M., Carpenter, J.~R., Morris, T.~P., Wood, A.~M., and Petersen, I.
  (2019).
\newblock Population-calibrated multiple imputation for a binary categorical
  covariate in categorical regression models.
\newblock {\em Statistics in Medicine\/}, {\bf 38}, 792--808.

\bibitem[Polson {\em et~al.}(2013)Polson, Scott, and Windle]{Polson2013}
Polson, N.~G., Scott, J.~G., and Windle, J. (2013).
\newblock Bayesian inference for logistic models using {P}\'{o}lya-–{G}amma
  latent variables.
\newblock {\em Journal of the American Statistical Association\/}, {\bf 108},
  1339--1349.

\bibitem[Raghunathan {\em et~al.}(2001)Raghunathan, Lepkowski, Hoewyk, and
  Solenberger]{Raghunathan2001}
Raghunathan, T.~E., Lepkowski, J.~M., Hoewyk, J.~V., and Solenberger, P.
  (2001).
\newblock A multivariate technique for multiply imputing missing values.
\newblock {\em Survey Methodology\/}, {\bf 27}, 85--95.

\bibitem[Reiter(2008)Reiter]{Jerry2008}
Reiter, J.~P. (2008).
\newblock Multiple imputation when records used for imputation are not used or
  disseminated for analysis.
\newblock {\em Biometrika\/}, {\bf 95}, 933--946.

\bibitem[Rubin(1976)Rubin]{Rubin1976}
Rubin, D.~B. (1976).
\newblock Inference and missing data.
\newblock {\em Biometrika\/}, {\bf 63}, 581--592.

\bibitem[Rubin(1987)Rubin]{Rubin1987}
Rubin, D.~B. (1987).
\newblock {\em Multiple Imputation For Nonresponse In Surveys\/}.
\newblock John Wiley \& Sons, New York.

\bibitem[Sadinle and Reiter(2017)Sadinle and Reiter]{Sadinle_Reiter2017}
Sadinle, M. and Reiter, J.~P. (2017).
\newblock Itemwise conditionally independent nonresponse modelling for
  incomplete multivariate data.
\newblock {\em Biometrika\/}, {\bf 104}, 207--220.

\bibitem[Sadinle and Reiter(2019)Sadinle and Reiter]{Sadinle2019}
Sadinle, M. and Reiter, J.~P. (2019).
\newblock Sequentially additive nonignorable missing data modelling using
  auxiliary marginal information.
\newblock {\em Biometrika\/}, {\bf 106}, 889--911.

\bibitem[Schifeling {\em et~al.}(2015)Schifeling, Cheng, Reiter, and
  Hillygus]{Schifeling:Reiter}
Schifeling, T., Cheng, C., Reiter, J.~P., and Hillygus, D.~S. (2015).
\newblock Accounting for nonignorable unit nonresponse and attrition in panel
  studies with refreshment samples.
\newblock {\em Journal of Survey Statistics and Methodology\/}, {\bf 3},
  265--295.

\bibitem[Si and Zhou(2021)Si and Zhou]{yajuanBayesraking}
Si, Y. and Zhou, P. (2021).
\newblock Bayes-raking: {B}ayesian finite population inference with known
  margins.
\newblock {\em Journal of Survey Statistics and Methodology\/}, {\bf 8},
  833--855.

\bibitem[Si {\em et~al.}(2016)Si, Reiter, and Hillygus]{yajuan:reiter:hilly}
Si, Y., Reiter, J.~P., and Hillygus, D.~S. (2016).
\newblock Bayesian latent pattern mixture models for handling attrition in
  panel studies with refreshment samples.
\newblock {\em Annals of Applied Statistics\/}, {\bf 10}, 118--143.

\bibitem[Silver {\em et~al.}(1996)Silver, Anderson, and
  Abramson.]{Silver_Anderson_Abramson_1986}
Silver, B.~D., Anderson, B.~A., and Abramson., P.~R. (1996).
\newblock Who overreports voting?
\newblock {\em The American Political Science Review\/}, {\bf 80}, 613–24.

\bibitem[Tang(2022)Tang]{TangDis}
Tang, J. (2022).
\newblock {\em Bayesian Models for Combining Information from Multiple
  Sources\/}.
\newblock Ph.D. thesis, Duke University.

\bibitem[Valliant {\em et~al.}(2013)Valliant, Dever, and Kreuter]{Valliant2013}
Valliant, R., Dever, J.~A., and Kreuter, F. (2013).
\newblock {\em Practical Tools for Designing and Weighting Sample Surveys\/}.
\newblock Springer-Verlag, New York.

\end{thebibliography}

\end{document}


\title{Supplemental material for ``Using auxiliary marginal distributions in imputations for nonresponse while accounting for survey weights, with application to estimating voter turnout''}
\author{Jiurui Tang, D. Sunshine Hillygus, Jerome P. Reiter}
\date{}
\maketitle


\section{Introduction}
In Section \ref{sim}, we present the results of simulation studies of the  hybrid missingness MD-AM model with the intercept matching algorithm.  To write the models, we use the notation described in the main text.  In Section \ref{CPSmodel}, we present the hybrid missingness MD-AM model without measurement error used for the turnout analysis in the main text.  In Section \ref{CPSmeasurementerror}, we present the additional steps for the intercept matching algorithm to account for measurement error modeling.  We also present results of the turnout analysis using the measurement error model layered on top of the  hybrid missingness MD-AM model.

\section{Simulation Studies}
\label{sim}
This section presents simulation studies that examine the repeated sampling performance of the hybrid missingness MD-AM modeling strategy and intercept matching algorithm. Section \ref{setup} describes eight simulation scenarios that we use for two situations, namely when design weights $w_i^d$ are available for unit respondents (presented in Section \ref{res_wd}) and when only adjusted weights $w_i^a$ are available for unit respondents (presented in Section \ref{res_wa}). 
\subsection{Simulation Set-up}
\label{setup}


In each 
simulation scenario, we construct a simulated population comprising $N = 133427$ individuals measured on two variables, $(X_1, X_2)$.  This $N$ matches the number of  non-zero household weights from the 2012 Current Population Survey data, abbreviated as 2012 CPS in the supplement---we emphasize that these are 2012 CPS data rather than the 2018 CPS data used in the turnout analysis in the main text---which we treat as design weights for purposes of random sampling.  We create each simulation population by sampling from the hybrid missingness MD-AM model in \eqref{sim_set_2.1}---\eqref{sim_set_2.4}. These use probit models to generate the data. We arbitrarily attach a design weight $w_i^d$ to each unit in the population.

\begin{align}
\mathbb{P}(U=1) &= \pi_U, \, \pi_U =  \Phi(\nu_0) \label{sim_set_2.1}\\
X_{2}|U &\sim Bernoulli(\pi_{x_{2}}),\, \pi_{x_{2}} = \Phi(\alpha_0 + \alpha_1 U) \label{sim_set_2.2}\\
X_{1}|X_{2},U &\sim Bernoulli(\pi_{x_{1}}),\, \pi_{x_{1}} = \Phi(\beta_0 + \beta_1 X_{2} + \beta_2 U) \label{sim_set_2.3}\\
\mathbb{P}(R_1^x=1|X_1,X_2,U) &= \pi_R, \, \pi_R =  \Phi(\gamma_0 + \gamma_1 X_2) \label{sim_set_2.4}.
\end{align}

In each simulation scenario, we set $\nu_0 = -1.2$, leading to approximately 11\% unit nonrespondents. This matches the percentage of households that do not respond in the 2012 CPS data. We set $(\gamma_0,\gamma_1) = (-1.05,0.2)$, as this setting gives approximately 18\% item nonresponse to match the item nonresponse rate of the vote variable in the 2012 CPS data. We set $(\alpha_0 = 0.3, \beta_0 = 0.2)$. We use two values for each parameter in $(\alpha_1, \beta_1, \beta_2)$, resulting in $2^3 = 8$ simulation scenarios in total. We expect  these three variables to affect the performance of the models. $\alpha_1$ represents the relationship between unit nonresponse $U$ and $X_2$. We pick $\alpha_1$ to be either $-0.5$ or $-2.0$, one for a weak relationship and the other for a strong relationship. $\beta_1$ represents the relationship between $X_1$ and $X_2$, for which we pick $0.4$ and $2.0$ to show weak and strong associations. $\beta_2$ represents the relationship between unit nonresponse $U$ and $X_1$. We pick $\beta_2$ to be either $-0.5$ or $-2.0$ as we do for $\alpha_1$.

Table \ref{meanX2} and Table \ref{meanX1} display the means of $X_1$ and $X_2$ for unit respondents and nonrespondents in each of the eight simulation settings. When $\alpha_1 = -2.0$, the difference in the mean of $X_2$ for unit respondents and nonrespondents is much larger than when $\alpha_1 = -0.5$. The difference in the mean of $X_1$ is more nuanced. When $(\alpha_1, \beta_1) = (-0.5,0.4)$, we see the least difference in the mean of $X_1$ across unit respondents and non-respondents, around $0.2$. When $(\alpha_1, \beta_1, \beta_2) = (-2.0,2.0,-2.0)$, we see the largest difference in the mean of $X_1$ across unit respondents and non-respondents, which are 0.833 and 0.062 respectively. This represents a very substantial nonresponse bias. When $(\alpha_1, \beta_2) = (-0.5, -2.0)$, the differences in the means of $X_1$ are also quite large, around $0.6$.

\begin{table}[t]
\centering
\caption{Means of $X_2$ in the eight simulation settings for unit respondents and nonrespondents.}
\begin{tabular}{|c|c|c|}
\hline
& $\alpha_1 = -0.5$ & $\alpha_1 = -2.0$ \\
\hline
$U=0$ & .618 & .618 \\
\hline
$U=1$ & .421 & .044 \\
\hline
\end{tabular}
\label{meanX2}
\end{table}

\begin{table}[t]
\centering
\caption{Means of $X_1$ in the eight simulation settings for unit respondents and nonrespondents.}
\begin{tabular}{|cc|c|c|c|c|}
\hline
 & \multirow{2}{*}{} & \multicolumn{2}{c}{$\alpha_1 = -0.5$} & \multicolumn{2}{|c|}{$\alpha_1 = -2.0$} \\
\cline{3-6}
 & & $\beta_1 = 0.4$ & $\beta_1 = 2.0$ & $\beta_1 = 0.4$ & $\beta_1 = 2.0$ \\
\hline 
\multirow{2}{*}{U = 0} & \multicolumn{1}{|l|}{$\beta_2 = -0.5$} & .668 & .830 & .671 & .833 \\
\cline{2-6}
& \multicolumn{1}{|l|}{$\beta_2 = -2.0$} & .670 & .830 & .672 & .833 \\
\hline
\multirow{2}{*}{U = 1} & \multicolumn{1}{|l|}{$\beta_2 = -0.5$} & .452 & .625 & .386 & .409 \\
\cline{2-6}
& \multicolumn{1}{|l|}{$\beta_2 = -2.0$} & .053 & .263 & .038 & .062 \\
\hline
\end{tabular}
\label{meanX1}
\end{table}

From each population, we independently sample 1000 data sets using Poisson sampling with inclusion probability $\pi_i  = 1/w_i^d$ for unit $i$. Then, we fit the hybrid missingness MD-AM model with the intercept matching algorithm as described in the main text, which is like \eqref{sim_set_2.1}---\eqref{sim_set_2.4} except using logistic regressions.  We run 10000 iterations, discard the first 5000 runs as burn-in, and keep 5000 posterior samples. We create $L=50$ multiple imputation data sets, $\pmb{Z} = (\pmb{Z}^{(1)},\dots,\pmb{Z}^{(50)})$, from every 100 posterior samples. For each completed data set $\pmb{Z}^{(\ell)}$, we compute the estimates of $T_{X1}, T_{X2}$, conditional margins and their standard errors with the ``survey'' package \citep{surveyR} in R. We compute the design-based estimates of $(\alpha_0, \alpha_1, \beta_0, \beta_1, \beta_2)$ along with the corresponding standard errors using the survey-weighted generalized linear models option in the ``survey'' package. We also compute estimates of $(\gamma_0, \gamma_1, \gamma_2)$, which do not depend on the design weights, along with the corresponding standard errors, using the standard generalized linear models routine in R. 

We also analyze the sampled data sets in each simulation setting without using the auxiliary margins, as specified in (9)---(12) of the main text. We fit the  same models for $X_1$ and $X_2$ for unit respondents and unit nonrespondents. There is no information available to adjust the intercept for unit nonrespondents.  We skip steps IM1, IM2, and IM4 from the main text, and draw imputations from $X_2|U=1 \sim Bernoulli\Big(1/\big(1+\exp(-\alpha_0^{(t+1)})\big)\Big)$ when sampling $X_2$ for unit nonrespondents. Similarly, we skip IM6---IM8, and draw imputations of $X_1$ for unit nonrespondents from $X_1|U=1 \sim Bernoulli\Big(1/\big(1+\exp(-\beta_0^{(t+1)}-\beta_1^{(t+1)}X_2)\big)\Big)$. We refer to analysis results without auxiliary margins as ICIN, as the nonresponse mechanism we specify under this scenario is itemwise conditionally independent nonresponse \citep{Sadinle_Reiter2017}.

\subsection{Simulations And Results When Design Weights Available}
\label{res_wd}
We begin with the case of available design weights for unit respondents.  We first look at results for survey variables, which are of primary interest in practice. Table \ref{DW_alphaWeak} displays results for the four simulation settings with $\alpha_1 = -0.5$, and Table \ref{DW_alphaStrong} displays the results with $\alpha_1 = -2.0$. For each simulation scenario, we summarize six quantities: estimates of $T_{X_1}$ and $T_{X_2}$ as well as the conditional probabilities of $X_1$ given $X_2$ and $X_2$ given $X_1$. The first columns of those tables are labelled ``truth'', representing the true population quantities in each simulation setting. CI coverage is the 95\% confidence interval (CI) empirical coverage rate  for each estimand across 1000 samples. ``SD'' is the standard deviation of the 1000 sample means for each estimand. ``Avg. est. SD" is the square root of the average of the 1000 estimated variances, computed based on the \citet{Rubin1987} multiple imputation combining rules.  ``Pre-SD'' stands for pre-missing standard deviation, the standard deviation of the 1000 point estimates calculated from the 1000 sampled full data sets with original weights. The $b_L$ is the between imputation standard deviation, part of Rubin's combining rule. The last column is obtained from taking the mean of $b_L/L$ for each estimand and then taking its square root, where $L = 50$ is the number of multiple imputation data sets. As we discuss, having $\sqrt{b_L/L}$ is useful for assessing the multiple imputation variance estimates. 

We begin by examining the results for $T_{X_1}$ and $T_{X_2}$. Point estimates from MD-AM are less biased than those under ICIN. This is not surprising, as the auxiliary margins provide extra information for the imputation modeling. We do see MD-AM overestimate $T_{X_2}$ in Table \ref{DW_alphaStrong}. As discussed in the main text, in these scenarios the mean of $X_2$ for unit respondents is much larger than that for unit non-respondents. Because of the large unit nonresponse bias, in the scenarios in Table \ref{DW_alphaStrong} the weighted sum of $X_2$ of unit respondents alone could be larger than $T_{X_2}$, making it not possible for the imputation of $X_2$ for unit nonrespondents to adjust the estimate down sufficiently. This causes an upward bias in the multiple imputation estimates of $T_{X_2}$.  This bias does not exist in Table \ref{DW_alphaWeak}, where the nonresponse bias is not so extreme. 

Across all eight simulation settings, the CI coverage rates for $T_{X_1}$ and $T_{X_2}$ are much higher under MD-AM than under ICIN. This happens because, first, MD-AM gives more accurate point estimates and second, the average estimated standard deviations based on Rubin's multiple imputation combining rules tend to be positively biased when using the margins for imputation. Assuming homogeneous distribution of $X_2$ across unit respondents and nonrespondents, as done under ICIN model specification, is least problematic when $\alpha_1 = -0.5$ though results for MD-AM still outperform those for ICIN. On the other hand, when we set $\alpha_1 = -2.0$, estimates under ICIN become so biased that all CI coverage rates of $T_{X_2}$ equal 0. We see similar patterns in the coverage rates for $T_{X_1}$.

When we use auxiliary information in generating multiple imputation data sets, the true standard deviation of the point estimates tend to be smaller than the average estimated standard deviation, especially when estimating quantities that employ auxiliary margins directly, such as $T_{X_1}$ and $T_{X_2}$. The estimated standard deviations from the multiple imputation combining rules are at least three times as large as the true standard deviations under MD-AM. 
The true standard deviations of the multiple imputation point estimators of $T_{X_1}$ and $T_{X_2}$ are even smaller than the pre-missing standard deviation. The MD-AM model creates completed data sets that attempt to match draws of the estimated totals, and the average of those estimated totals converges to $T_{X_1}$ and $T_{X_2}$ due to the law of large numbers. The true standard deviations are in general between $\sqrt{b_L/L}$ and the average estimated standard deviation from the multiple combining rules. When design weights are known for unit respondents, we suggest using standard deviations calculated from Rubin's multiple imputation combining rules, even though these may be conservative, because $\sqrt{b_L/L}$ underestimates the uncertainty significantly when nonresponse bias is large or correlation between $X_1$ and $X_2$ is strong. The true standard deviations under the ICIN model are larger than those under the MD-AM model due to the employment of auxiliary margins in the latter.

Next, we examine estimates of conditional distributions. When design weights are known for unit respondents, estimates of conditional distributions obtained from MD-AM generally are more accurate than those under ICIN without using margins. There are some instances when the two models achieve similar level of accuracy for one or two of the four conditionals, for example in Table \ref{DW_alphaStrong} simulation (d), but the MD-AM model always performs better when we consider all four conditional margins as a whole. The CI coverage rates under MD-AM are always higher than those under ICIN due to more accurate point estimates and larger average estimated standard deviations, as we have seen when analyzing results for $T_{X_1}$ and $T_{X_2}$. 

The average estimated multiple imputation standard deviations of the conditional margins under MD-AM are larger than the corresponding true standard deviations, which may or may not be smaller than the pre-missing standard deviations. 
The true standard deviations of the conditional margins are again in between the average estimated standard deviations and $\sqrt{b_L/L}$. For conditional margins under MD-AM, we  suggest using Rubin's multiple imputation combining rules. Average estimated standard deviations of conditionals under ICIN are similar to the true standard deviations, which are also close to the pre-missing standard deviations.  

Although arguably of less interest than the totals and conditional means, simulation results for model parameters with known design weights for unit respondents are shown in Table \ref{DW_alphaWeak_par} and \ref{DW_alphaStrong_par}. The columns labelled ``truth''  are obtained by averaging the estimated model parameters in 1000 pre-missing samples. Estimates of $(\alpha_0, \gamma_0, \gamma_1)$ are similar and close to the truth values under MD-AM and ICIN, as these three parameters are estimable with the observed data alone. $\alpha_1$ and $\beta_2$ are the most difficult parameters to estimate, as the only information on them derives from the  auxiliary margins. Estimates of $\alpha_1$ and $\beta_2$ under ICIN are close to zero in all simulation settings, as those two parameters are not identifiable without using margins and our prior distributions for them are centered at zero. Estimates of $\alpha_1$ under MD-AM are more accurate when the unit nonresponse bias in $X_2$ is generated by $\alpha_1 = -0.5$ than when the unit nonresponse bias in $X_2$ is generated by $\alpha_1 = -2.0$. Under the same correlation between $X_1$ and $X_2$, estimates of $\beta_2$ under MD-AM are more accurate when the unit nonresponse bias in $X_1$ is weak than when the unit nonresponse bias in $X_1$ is strong, except for Table \ref{DW_alphaWeak_par} simulation (c) and (d). Average estimated standard deviations under MD-AM and ICIN are similar except for $\alpha_1$ and $\beta_2$, because (except for $\alpha_1$ and $\beta_2$), estimating these model parameters does not directly involve auxiliary margins. 

\begin{landscape}
\begin{table}[h!]
\footnotesize
\centering
\caption[Simulation results for totals and conditional margins assuming design weights for unit respondents are known and $\alpha_1 = -0.5$.]{Simulation results for totals and conditional margins assuming design weights for unit respondents are known and $\alpha_0 = 0.3, \alpha_1 = -0.5, \beta_0 = 0.2, \nu_0 = -1.2, \gamma_0 = -1.05, \gamma_1 = 0.2$. Results averaged over 1000 sampled data sets from the simulated population, using Poisson sampling and 50 multiple imputed data sets.}
\begin{tabular}{l|ccc|cc|ccc|cc|c}
\hline
 & \multicolumn{3}{c|}{Estimate} & \multicolumn{2}{c|}{CI Coverage \%} & \multicolumn{3}{c|}{SD} & \multicolumn{2}{c|}{Avg. Est. SD} & 
$\sqrt{b_m/m}$ \\
 & Truth & MDAM & ICIN & MDAM & ICIN & Pre-SD & MDAM & ICIN & MDAM & ICIN & MDAM \\
 \hline
 \multicolumn{12}{l}{\textbf{(a)} $\pmb{\alpha_1 = -0.5, \beta_1 = 0.4, \beta_2 = -0.5}$} \\
\hspace{4mm}$T_{X_2}$ &  79477 & 79460 & 82450 & 100  & 36.2  & 1348.2 & 189.9 & 848.0 & 1976.7 & 1326.6 & 188.1 \\
\hspace{4mm}$T_{X_1}$ & 85981 & 85985 & 89235 & 100  & 32.1  & 1331.3 & 206.6 & 828.3 & 1864.3 & 1425.9 & 202.8 \\
\hspace{4mm}$X_2=0|X_1=0$ & .514 & .511 & .483 & 100  & 31.8  & .011 & .010 & .011 & .019 & .012 & .002  \\
\hspace{4mm}$X_2=0|X_1=1$ & .344 & .345 & .332 & 100  &  67.7  & .008 & .005 & .008 & .012 & .008 & .001 \\
\hspace{4mm}$X_1=0|X_2=0$ & .452 & .450 & .419 & 100  & 16.4  & .010 & .009 & .010 & .019 & .011 & .002 \\
\hspace{4mm}$X_1=0|X_2=1$ & .290 & .292 & .277 & 100  & 63.7  & .007 & .006  & .008 & .012 & .008 & .001 \\
\multicolumn{12}{l}{\textbf{(b)} $\pmb{\alpha_1 = -0.5, \beta_1 = 0.4, \beta_2 = -2.0}$} \\
\hspace{4mm}$T_{X_2}$ &  79499 & 79493 & 82545 & 100  & 31.7  & 1332.8  & 184.4 & 814.6 & 1853.3 & 1326.9 & 185.9 \\
\hspace{4mm}$T_{X_1}$ & 79834 & 80248 & 89128 & 99.3  & 0  & 1333.8 & 495.0 & 819.1 & 1732.3 & 1424.5 & 157.3 \\
\hspace{4mm}$X_2=0|X_1=0$ & .515 & .517 & .489 & 100  & 42.3  & .010 & .012 & .011 & .025 & .012 & .003 \\
\hspace{4mm}$X_2=0|X_1=1$ & .329 & .330 & .328 & 98.7  & 96.7  & .008 & .007 & .007 & .009 & .008 & $<.001$ \\
\hspace{4mm}$X_1=0|X_2=0$ & .512 & .509 & .426 & 99.5  & 0  & .010 & .011 & .010 & .017 & .011 & .002 \\
\hspace{4mm}$X_1=0|X_2=1$ & .327 & .323 & .274 & 99.1  & 0  & .008 & .009 & .007 & .015 & .008 & .002 \\
\multicolumn{12}{l}{\textbf{(c)} $\pmb{\alpha_1 = -0.5, \beta_1 = 2.0, \beta_2 = -0.5}$} \\
\hspace{4mm}$T_{X_2}$ &  79026 & 79021 &  82078 & 100  & 31  & 1283.6 & 181.9 & 822.4 & 1863.2 & 1321.9 & 188.5 \\
\hspace{4mm}$T_{X_1}$ & 107451 & 106255 & 110517 & 100  & 50.5  & 1527.5 & 364.5 & 676.0 & 1838.9 & 1540.0 & 149.9 \\
\hspace{4mm}$X_2=0|X_1=0$ & .948 & .946 & .949 & 99.7  & 96.4  & .006 & .008 & .007 & .013 & .008 & .002 \\
\hspace{4mm}$X_2=0|X_1=1$ & .277 & .270 & .268 & 100  & 73.3  & .006 & .003 & .006 & .013 & .007 & .002 \\
\hspace{4mm}$X_1=0|X_2=0$ & .453 & .472 & .424 & 98.9  & 28.9  & .010 & .008 & .011 & .020 & .011 & .002 \\
\hspace{4mm}$X_1=0|X_2=1$ & .017 & .019 & .014 & 99.9  & 75.5  & .002 & .003 & .002 & .005 & .002 & .001 \\
\multicolumn{12}{l}{\textbf{(d)} $\pmb{\alpha_1 = -0.5, \beta_1 = 2.0, \beta_2 = -2.0}$} \\
\hspace{4mm}$T_{X_2}$ &  79592 & 79577 & 82583 & 100  & 33.7  & 1272.5 & 186.1 & 804.6 & 1873.8 & 1328.3 & 189.7 \\
\hspace{4mm}$T_{X_1}$ & 102095 & 103364 & 111004 & 100  & 0  & 1441.6 & 254.0 & 635.6 & 1948.9 & 1543.2 & 176.7 \\
\hspace{4mm}$X_2=0|X_1=0$ & .874 & .900 & .946 & 96  & 0  & .008 & .017 & .007 & .039 & .008 & .005 \\
\hspace{4mm}$X_2=0|X_1=1$ & .259 & .259 & .267 & 99.7  & 82  & .006 & .005 & .006 & .008 & .007 & .005 \\
\hspace{4mm}$X_1=0|X_2=0$ & .509 & .502 & .417 & 98.3  & 0  & .008 & .010 & .010 & .015 & .011 & .001 \\
\hspace{4mm}$X_1=0|X_2=1$ & .050 & .038 & .015 & 93.7  & 0  & .003 & .006 & .002 & .016 & .002 & .002 \\
\hline
\end{tabular}
\label{DW_alphaWeak}
\end{table}
\end{landscape}

\begin{landscape}
\begin{table}[h!]
\footnotesize
\centering
\caption[Simulation results for totals and conditional margins assuming design weights for unit respondents are known and $\alpha_1 = -2.0$.]{Simulation results for totals and conditional margins assuming design weights for unit respondents are known and $\alpha_0 = 0.3, \alpha_1 = -2.0, \beta_0 = 0.2, \nu_0 = -1.2, \gamma_0 = -1.05, \gamma_1 = 0.2$. Results averaged over 1000 sampled data sets from the simulated population, using Poisson sampling and 50 multiple imputed data sets.}
\begin{tabular}{l|ccc|cc|ccc|cc|c}
\hline
 & \multicolumn{3}{c|}{Estimate} & \multicolumn{2}{c|}{CI Coverage \%} & \multicolumn{3}{c|}{SD} & \multicolumn{2}{c|}{Avg. Est. SD} & 
$\sqrt{b_m/m}$ \\
 & Truth & MDAM & ICIN & MDAM & ICIN & Pre-SD & MDAM & ICIN & MDAM & ICIN & MDAM \\
 \hline
 \multicolumn{12}{l}{\textbf{(a)} $\pmb{\alpha_1 = -2.0, \beta_1 = 0.4, \beta_2 = -0.5}$} \\
\hspace{4mm}$T_{X_2}$ &  73451 & 73879 & 82167 & 99.3  & 0  & 1249.1 & 520.9 & 819.3 & 1582.5 & 1324.3 & 132.8 \\
\hspace{4mm}$T_{X_1}$ & 84985 & 84983 & 89221 & 100  & 4.4  & 1333.2 & 203.0 & 819.7 & 1965.3 & 1424.4 & 201.3 \\
\hspace{4mm}$X_2=0|X_1=0$ & .581  & .577 & .491 & 99.7  & 0  & .010 & .012 & .011 & .018 & .012 & .002 \\
\hspace{4mm}$X_2=0|X_1=1$ & .375  & .371 & .331 & 98.9  & 0  & .007 & .009 & .007 & .014 & .008 & .002 \\
\hspace{4mm}$X_1=0|X_2=0$ & .469  & .470 & .423 & 100  & 1  & .009 & .010 & .010 & .024 & .011 & .003 \\
\hspace{4mm}$X_1=0|X_2=1$ & .277 & .277 & .274 & 98.5  & 97.1   & .008 & .008 & .008 & .009 & .008 & .001 \\
\multicolumn{12}{l}{\textbf{(b)} $\pmb{\alpha_1 = -2.0, \beta_1 = 0.4, \beta_2 = -2.0}$} \\
\hspace{4mm}$T_{X_2}$ & 73410 & 73847 & 82216 & 99.1  & 0  & 1284.7  & 546.2 & 829.5 & 1580.7 & 1325.4 & 132.5 \\
\hspace{4mm}$T_{X_1}$ & 79506 & 79995 & 89183 & 99.6 & 0  & 1306.0 & 570.4 & 821.4 & 1686.4 & 1425.3 & 147.3 \\
\hspace{4mm}$X_2=0|X_1=0$ & .617 & .608 & .488 & 98.8  & 0  & .010 & .011 & .011 & .019 & .012 & .002 \\
\hspace{4mm}$X_2=0|X_1=1$ & .336 & .339 & .332 & 97.4  & 93.8  & .008 & .009 & .008 & .011 & .008 & .001 \\
\hspace{4mm}$X_1=0|X_2=0$ & .555 & .544 & .422 & 97.9  & 0  & .010 & .010 & .011 & .019 & .011 & .002 \\
\hspace{4mm}$X_1=0|X_2=1$ & .281 & .284 & .276 & 98.7  & 92.4  & .008 & .009 & .008 & .012 & .008 & .001 \\
\multicolumn{12}{l}{\textbf{(c)} $\pmb{\alpha_1 = -2.0, \beta_1 = 2.0, \beta_2 = -0.5}$} \\
\hspace{4mm}$T_{X_2}$ &  73808 & 74258 & 82693 & 99.2  & 0  & 1285.5  & 550.7 & 832.5 & 1571.9 & 1327.0 & 130.0 \\
\hspace{4mm}$T_{X_1}$ & 104446 & 104198 & 110972 & 100  & 0  & 1512.8 & 304.5 & 655.0 & 2054 & 1542.6 & 198.2  \\
\hspace{4mm}$X_2=0|X_1=0$ & .965 & .964 & .949 & 98.2  & 54.5  & .005 & .005 & .007 & .006 & .008 & .001 \\
\hspace{4mm}$X_2=0|X_1=1$ & .303 & .297 & .265 & 99.2  & 0  & .007 & .005 & .007 & .015 & .007 & .002 \\
\hspace{4mm}$X_1=0|X_2=0$ & .469 & .476 & .420 & 100  & 1  & .010 & .005 & .011 & .027 & .011 & .004 \\
\hspace{4mm}$X_1=0|X_2=1$ & .014 & .014 & .014 & 98  & 96.7  & .002 & .002 & .002 & .002 & .002 & $<.001$ \\
\multicolumn{12}{l}{\textbf{(d)} $\pmb{\alpha_1 = -2.0, \beta_1 = 2.0, \beta_2 = -2.0}$} \\
\hspace{4mm}$T_{X_2}$ &  73582 & 73986 & 82344 & 99.7  & 0  & 1208.3  & 501.8 & 854.8 & 1578.3 & 1324.3 & 132.0 \\
\hspace{4mm}$T_{X_1}$ & 98814 & 99482 & 110485 & 99.7  & 0  & 1373.6 & 457.9 & 649.9 & 1883.2 & 1539.0 & 165.2 \\
\hspace{4mm}$X_2=0|X_1=0$ & .961 & .957 & .948 & 99.6  & 68.3  & .005 & .008 & .007 & .018 & .008 & .002 \\
\hspace{4mm}$X_2=0|X_1=1$ & .269 & .271 & .266 & 99.1  & 91.4  & .007 & .006 & .007 & .010 & .007 & .001 \\
\hspace{4mm}$X_1=0|X_2=0$ & .556 & .546 & .426 & 97.9  & 0  & .010 & .008 & .011 & .018 & .011 & .002 \\
\hspace{4mm}$X_1=0|X_2=1$ & .018 & .020 & .015 & 98.9  & 60  & .002  & .004 & .002 & .009 & .002 & .001 \\
\hline
\end{tabular}
\label{DW_alphaStrong}
\end{table}
\end{landscape}

\begin{table}[h!]
\footnotesize
\centering
\caption[Simulation results for model parameters assuming design weights for unit respondents are known and $\alpha_1 = -0.5$.]{Simulation results for model parameters assuming design weights for unit respondents are known and $\alpha_0 = 0.3, \alpha_1 = -0.5, \beta_0 = 0.2, \nu_0 = -1.2, \gamma_0 = -1.05, \gamma_1 = 0.2$. Results averaged over 1000 sampled data sets from the simulated population, using Poisson sampling and 50 multiple imputed data sets.}
\begin{tabular}{l|ccc|cc|ccc|cc}
\hline
 & \multicolumn{3}{c|}{Estimate} & \multicolumn{2}{c|}{CI Coverage \%} & \multicolumn{3}{c|}{SD} & \multicolumn{2}{c}{Avg. Est. SD} \\
 & Truth & MDAM & ICIN & MDAM & ICIN & Pre-SD & MDAM & ICIN & MDAM & ICIN \\
 \multicolumn{11}{l}{\textbf{(a)} $\pmb{\alpha_1 = -0.5, \beta_1 = 0.4, \beta_2 = -0.5}$} \\
\hspace{4mm}$\alpha_0$ & .300 & .300 & .300 & 95.5  & 95.5  & .018 & .018 & .018  & .018 & .018 \\
\hspace{4mm}$\alpha_1$ & -.506 & -.513 & -.002 & 99.7  & 0  & .037 & .178 & .012 & .256 & .057 \\
\hspace{4mm}$\beta_0$ & .202 & .205 & .205 & 97  & 96.7  & .026 & .027 & .027 & .030 & .030 \\
\hspace{4mm}$\beta_1$ & .396 & .388 & .388 & 97.1  & 97.7  & .033 & .035 & .035 & .038 & .038 \\
\hspace{4mm}$\beta_2$ & -.498 & -.493 & -.004 & 99.8  & 0  & .039 & .167 & .012 & .268 & .058 \\
\hspace{4mm}$\gamma_0$ & -1.06 & -1.06 & -1.06 & 96  & 96  & .025 & .025 & .025 & .026 & .026 \\
\hspace{4mm}$\gamma_1$ & .210 & .210 & .210 & 96.6  & 96.6  & .030 & .030 & .030 & .032 & .032 \\
 \multicolumn{11}{l}{\textbf{(b)} $\pmb{\alpha_1 = -0.5, \beta_1 = 0.4, \beta_2 = -2.0}$} \\
\hspace{4mm}$\alpha_0$ & .302 & .302 & .302 & 96.5  & 96.5  & .017 & .017 & .017  & .018 & .018 \\
\hspace{4mm}$\alpha_1$ & -.513 & -.522 & -.001 & 99.6  & 0  & .037 & .176 & .012 & .259 & .057 \\
\hspace{4mm}$\beta_0$ & .195 & .189 & .187 & 97.3  & 96.4  & .026 & .027 & .027 & .030 & .030 \\
\hspace{4mm}$\beta_1$ & .405 & .410 & .413 & 97.3  & 96.9  & .034 & .035 & .035 & .040 & .038 \\
\hspace{4mm}$\beta_2$ & -1.998 & -3.055 & -.002 & 90.9  & 0  & .058 & 1.255 & .012 & 1.673 & .058 \\
\hspace{4mm}$\gamma_0$ & -1.070 & -1.070 & -1.070 & 97.3  & 97.3  & .024 & .024 & .024 & .026 & .026 \\
\hspace{4mm}$\gamma_1$ & .221 & .221 & .221 & 97.4  & 97.4  & .028 & .028 & .028 & .032 & .032 \\
 \multicolumn{11}{l}{\textbf{(c)} $\pmb{\alpha_1 = -0.5, \beta_1 = 2.0, \beta_2 = -0.5}$} \\
\hspace{4mm}$\alpha_0$ & .292 & .292 & .292 & 95.9  & 95.9  & .017 & .017 & .017  & .018 & .018 \\
\hspace{4mm}$\alpha_1$ & -.498 & -.516 & .003 & 99.7  & 0  & .037 & .170 & .013 & .252 & .056 \\
\hspace{4mm}$\beta_0$ & .198 & .185 & .194 & 95.8  & 95.9  & .028 & .028 & .029 & .030 & .030 \\
\hspace{4mm}$\beta_1$ & 1.993 & 2.049 & 2.002 & 91  & 96  & .057 & .067 & .065 & .072 & .068 \\
\hspace{4mm}$\beta_2$ & -.477 & -.782 & -.004 & 97.4  & 0  & .047 & .128 & .017 & .332 & .081 \\
\hspace{4mm}$\gamma_0$ & -1.026 & -1.026 & -1.026 & 96.6  & 96.6  & .024 & .024 & .024 & .025 & .025 \\
\hspace{4mm}$\gamma_1$ & .174 & .174 & .174 & 95.8  & 95.8  & .030 & .030 & .030 & .032 & .032 \\
 \multicolumn{11}{l}{\textbf{(d)} $\pmb{\alpha_1 = -0.5, \beta_1 = 2.0, \beta_2 = -2.0}$} \\
\hspace{4mm}$\alpha_0$ & .303 & .303 & .303 & 96.8  & 96.8  & .017 & .017 & .017  & .018 & .018 \\
\hspace{4mm}$\alpha_1$ & -.500 & -.510 & -.002 & 99.8  & 0  & .038 & .163 & .012 & .249 & .057 \\
\hspace{4mm}$\beta_0$ & .204 & .199 & .208 & 96.5  & 97.1  & .025 & .028 & .027 & .030 & .030 \\
\hspace{4mm}$\beta_1$ & 1.982 & 2.024 & 1.977 & 94.3  & 98  & .056 & .067 & .062 & .072 & .068 \\
\hspace{4mm}$\beta_2$ & -2.035 & -1.847 & .006 & 84.4  & 0  & .061 & .507 & .016 & .975 & .082 \\
\hspace{4mm}$\gamma_0$ & -1.046 & -1.046 & -1.046 & 97.6  & 97.6  & .023 & .023 & .023 & .026 & .026 \\
\hspace{4mm}$\gamma_1$ & .191 & .191 & .191 & 97.1  & 97.1  & .029 & .029 & .029 & .032 & .032 \\
\hline
\end{tabular}
\label{DW_alphaWeak_par}
\end{table}

\begin{table}[h!]
\footnotesize
\centering
\caption[Simulation results for model parameters assuming design weights for unit respondents are known and $\alpha_1 = -2.0$.]{Simulation results for model parameters assuming design weights for unit respondents are known and $\alpha_0 = 0.3, \alpha_1 = -2.0, \beta_0 = 0.2, \nu_0 = -1.2, \gamma_0 = -1.05, \gamma_1 = 0.2$. Results averaged over 1000 sampled data sets from the simulated population, using Poisson sampling and 50 multiple imputed data sets.}
\begin{tabular}{l|ccc|cc|ccc|cc}
\hline
 & \multicolumn{3}{c|}{Estimate} & \multicolumn{2}{c|}{CI Coverage \%} & \multicolumn{3}{c|}{SD} & \multicolumn{2}{c}{Avg. Est. SD} \\
 & Truth & MDAM & ICIN & MDAM & ICIN & Pre-SD & MDAM & ICIN & MDAM & ICIN \\
 \multicolumn{11}{l}{\textbf{(a)} $\pmb{\alpha_1 = -2.0, \beta_1 = 0.4, \beta_2 = -0.5}$} \\
\hspace{4mm}$\alpha_0$ & .295 & .295 & .295 & 95.6  & 95.6  & .017 & .017 & .017  & .018 & .018 \\
\hspace{4mm}$\alpha_1$ & -2.027 & -3.181 & -.002 & 88.8  & 0  & .063 & 1.255 & .012 & 1.649 & .056 \\
\hspace{4mm}$\beta_0$ & .195 & .193 & .193 & 98.2  & 97.8  & .026 & .027 & .026 & .030 & .030 \\
\hspace{4mm}$\beta_1$ & .402 & .406 & .406 & 97.2  & 97.2  & .035 & .036 & .035 & .040 & .038 \\
\hspace{4mm}$\beta_2$ & -.491 & -.520 & .008 & 99.8  & 0  & .041 & .187 & .011 & .307 & .058 \\
\hspace{4mm}$\gamma_0$ & -1.061 & -1.061 & -1.061 & 97.3  & 97.3  & .023 & .023 & .023 & .026 & .026 \\
\hspace{4mm}$\gamma_1$ & .216 & .216 & .216 & 98  & 97.7  & .028 & .028 & .028 & .032 & .032 \\
 \multicolumn{11}{l}{\textbf{(b)} $\pmb{\alpha_1 = -2.0, \beta_1 = 0.4, \beta_2 = -2.0}$} \\
\hspace{4mm}$\alpha_0$ & .296 & .296 & .296 & 96.9  & 96.9  & .017 & .018 & .017  & .018 & .018 \\
\hspace{4mm}$\alpha_1$ & -2.008 & -3.164 & -.005 & 88.4  & 0 & .061 & 1.255 & .012 & 1.645 & .057 \\
\hspace{4mm}$\beta_0$ & .196 & .198 & .198 & 96.1  & 96.1  & .028 & .028 & .028 & .030 & .030 \\
\hspace{4mm}$\beta_1$ & .401 & .398 & .399 & 96.7  & 96.8  & .037 & .037 & .036 & .040 & .038 \\
\hspace{4mm}$\beta_2$ & -1.965 & -3.168 & .001 & 89  & 0  & .064 & 1.253 & .012 & 1.692 & .058 \\
\hspace{4mm}$\gamma_0$ & -1.073 & -1.073 & -1.073 & 97  & 97  & .024 & .024 & .024 & .026 & .026 \\
\hspace{4mm}$\gamma_1$ & .224 & .224 & .224 & 96.8  & 96.9  & .029 & .029 & .029 & .032 & .032 \\
 \multicolumn{11}{l}{\textbf{(c)} $\pmb{\alpha_1 = -2.0, \beta_1 = 2.0, \beta_2 = -0.5}$} \\
\hspace{4mm}$\alpha_0$ & .305 & .305 & .305 & 94.7  & 94.7  & .017 & .017 & .017  & .018 & .018 \\
\hspace{4mm}$\alpha_1$ & -2.050 & -3.259 & .003 & 87.3  & 0  & .063 & 1.285 & .012 & 1.646 & .057 \\
\hspace{4mm}$\beta_0$ & .198 & .200 & .201 & 98  & 96.9  & .029 & .029 & .029 & .031 & .030 \\
\hspace{4mm}$\beta_1$ & 2.012 & 2.018 & 2.008 & 97  & 97.1 & .063 & .067 & .064 & .072 & .069 \\
\hspace{4mm}$\beta_2$ & -.486 & -.633 & .001 & 100  & 0  & .045 & .195 & .018 & .391 & .082 \\
\hspace{4mm}$\gamma_0$ & -1.059 & -1.059 & -1.059 & 97.7  & 97.7  & .024 & .024 & .024 & .026 & .026 \\
\hspace{4mm}$\gamma_1$ & .212 & .212 & .212 & 97.2  & 97.2  & .030 & .030 & .030 & .032 & .032 \\
 \multicolumn{11}{l}{\textbf{(d)} $\pmb{\alpha_1 = -2.0, \beta_1 = 2.0, \beta_2 = -2.0}$} \\
\hspace{4mm}$\alpha_0$ & .298 & .298 & .298 & 94.8  & 94.8  & .018 & .018 & .018  & .018 & .018 \\
\hspace{4mm}$\alpha_1$ & -2.043 & -3.173 & .004 & 88.7  & 0  & .062 & 1.247 & .013 & 1.648 & .057 \\
\hspace{4mm}$\beta_0$ & .193 & .186 & .188 & 95.7  & 96.3  & .028 & .029 & .027 & .031 & .030 \\
\hspace{4mm}$\beta_1$ & 1.983 & 2.009 & 1.998 & 96.4  & 97.2  & .059 & .065 & .061 & .072 & .068 \\
\hspace{4mm}$\beta_2$ & -2.048 & -3.164 & -.001 & 92.1  & 0  & .067 & 1.274 & .017 & 2.034 & .081 \\
\hspace{4mm}$\gamma_0$ & -1.033 & -1.033 & -1.033 & 97.9  & 97.9  & .024 & .024 & .024 & .026 & .026 \\
\hspace{4mm}$\gamma_1$ & .192 & .192 & .192 & 96.7  & 96.7 & .029 & .029 & .029 & .032 & .032 \\
\hline
\end{tabular}
\label{DW_alphaStrong_par}
\end{table}

\clearpage
\subsection{Simulations And Results With Adjusted Weights}
\label{res_wa}
In this section, we investigate the hybrid missingness MD-AM model with the intercept matching algorithm when we do not know the design weights for unit respondents. We use the same simulation set-up as in section \ref{res_wd}.  However, in these simulations, we need to mimic a nonresponse weighting adjustment to the unit respondents' design weights.   We employ a weighting class adjustment, in which we (i) divide the sampled data  into groups where unit respondents and nonrespondents are similar and (ii) increase the weights of unit respondents within each class to represent themselves as well as the unit nonrespondents' share of the population  \citep{Lohr2010}.  
Specifically, in each simulation run, we create two weighting adjustment classes based on the value of $X_1$. We estimate the response probability for each class by 
\begin{align}
\hat{\phi}_c = \frac{\sum_{i=1}^n w_i^dI(X_{i1} = c, U_i = 0)}{\sum_{i=1}^n w_i^dI(X_{i1} = c)},
\end{align}
where $c = \{0,1\}$. This is the ratio of the sum of weights for unit respondents over the sum of weights for all sampled units in class $c$. To obtain adjusted weights $w_i^a$ for unit respondents, we multiply each $W_i^d$ by $1/\hat{\phi}_c$ according to the $X_{i1}$ value, so that  
\begin{equation}
w_i^a =  \left\{\begin{array}{lr}
        w_i^d/\hat{\phi}_0, & \text{if } X_{i1} = 0\\
        w_i^d/\hat{\phi}_1, & \text{if } X_{i1} = 1
        \end{array}\right. \label{class_adj_eq}
\end{equation}

To create analysis weights $w_i^*$, we follow (19) in the main text. We note that this attempt to ``unadjust'' the $w_i^a$ does not get back to the corresponding $w_i^d$. We expect this might create potential biases; this is a consequence of not knowing the design weights. Nonetheless, absent any information about the unit nonrespondents' weights or population totals for the weighting classes, \eqref{class_adj_eq} can be viewed as a default procedure. Thus, it is worth evaluating the performance of the MD-AM model under this procedure. 

Before presenting the simulation results, we support a statement in section 3.1.3 of the main text. When design weights are unknown for unit respondents, whether we create weights for unit nonrespondents so that all the weights sum to the population size $N$ or its estimate $\hat{N}$ generally does not make much difference in inferences. Table \ref{pop1_N_W} displays results of 100 simulation runs with parameter settings at $(\alpha_1, \beta_1, \beta_2) =(-0.5,0.4, -0.5)$ when we use $N$ to create weights and 1000 simulation runs when we use $\hat{N}$ to create weights. We barely observe any difference in the results, especially for the survey variables.  Hence, we use $\hat{N}$ to create analysis weights for unit nonrespondents in the simulation studies in this section and in the 2018 CPS analysis in the main text.   

Table \ref{alphaWeak} and Table \ref{alphaStrong} display the simulation results when design weights for unit respondents are unavailable for analysis. Table \ref{alphaWeak} displays the results in the four simulation settings with $\alpha_1 = -0.5$, and Table \ref{alphaStrong} presents the results when $\alpha_1 = -2.0$. Point estimates of totals under MD-AM are less biased than those under ICIN, except for $T_{X2}$ in Table \ref{alphaWeak} simulation (c). In this case with $(\alpha_1,\beta_1,\gamma_2) = (-0.5,2.0,-0.5)$, nonresponse biases in $X_1$ and $X_2$ are small, and the correlation between $X_1$ and $X_2$ is strong. When imputing $X_1$ for unit nonrespondents, the correlation between $X_1$ and $X_2$ has a stronger effect on imputations than the nonresponse bias in $X_1$, resulting in less ones imputed for $X_1$ and hence an underestimate of $T_{X_1}$ in MD-AM. Also, because the nonresponse biases in $X_1$ and $X_2$ are modest, ignoring them under ICIN is not particularly problematic, so that the estimate of $T_{X_1}$ happens to be closer to the truth in this setting. Comparing Table \ref{DW_alphaStrong} and Table \ref{alphaStrong}, we also notice that estimates of $T_{X_2}$ from MD-AM are closer to the truth when the nonresponse bias in $X_2$ is large. We down-weight the unit respondents' adjusted weights in the process of creating weights for unit nonrespondents. This weight adjustment lowers the weighted sum of $X_2$ for unit respondents, resulting in more accurate estimates of $T_{X_2}$. 

Across all eight simulation settings, the CI coverage rates for $T_{X_1}$ and $T_{X_2}$ are much higher under MD-AM than under ICIN, mirroring what we have seen when the design weights are known for unit respondents. Estimates of the standard deviations from multiple imputation combining rules are almost ten times as large as the true standard deviations under MD-AM model specification. When design weights are unknown for unit respondents, we suggest using $\sqrt{b_L/L}$ for the estimated standard deviations of the estimated totals for variables with margins. Indeed, $\sqrt{b_L/L}$ for $T_{X_1}$ and $T_{X_2}$ are much closer to the corresponding true values than the average estimated standard deviations.

Next, we examine estimates of conditional margins. When design weights for unit respondents are not available, the performance of MD-AM in estimating conditional distributions is worse than when design weights for unit respondents are known.  However, it remains better than the performance of ICIN, except for the simulation setting in Table \ref{alphaStrong} simulation (b). In this setting, the relationships between $U$ and $X_1$ as well as $U$ and $X_2$ are both strong, whereas the relationship between $X_1$ and $X_2$ is weak. It is the most difficult simulation scenario to get  accurate estimates for the conditional margins. When the correlation between $U$ and $X_1$ is weaker at $\beta_2 = -0.5$, the CI coverage of the conditional margins under MD-AM is higher than when the correlation between $U$ and $X_1$ is at $\beta_2 = -2.0$, except for Table \ref{alphaWeak} simulations (a) and (b), where the performances under these two setting are similar, the latter slightly better. Stronger association between $X_1$ and $X_2$ leads to higher CI coverage in Table \ref{alphaStrong}, where $\alpha_1 = -2.0$. We do not observe similar advantages of a strong relationship between $X_1$ and $X_2$ when $\alpha_1 = -0.5$. 

Comparing Table \ref{DW_alphaWeak} and Table \ref{DW_alphaStrong} to Table \ref{alphaWeak} and Table \ref{alphaStrong}, the MD-AM model performs better when the design weights for unit respondents are available, especially in estimating conditional margins. The most obvious example is the comparison of Table \ref{DW_alphaStrong} simulation (b) to Table \ref{alphaStrong} simulation (b). This suggests that agencies can help researchers by publishing the design weights of the unit nonrespondents, in addition to any adjusted weights. 

We conclude this section on simulation results with a discussion of the model parameters. Table \ref{parameter} displays the results for the model parameters in one simulation setting. The overall patterns in the simulation results for model parameters are  similar in the other simulation scenarios. Recall that we do not know the weights for unit nonrespondents and have to do weight adjustment. This adjustment makes studying the quality of survey weighted estimates, namely $\pmb{\alpha}$ and $\pmb{\beta}$, challenging. The column labelled  ``Truth"  records the parameter values used in the data generating process, but they are not a good standard for comparison as we adjust the weights in every simulation run. Thus, our primary focus in Table \ref{parameter} is not to study how well we estimate the data generating parameters, but to discuss a few interesting points. 

Compared to when we know design weights for unit respondents, $\alpha_1$ and $\beta_2$ under MD-AM are more biased, especially $\beta_2$.  After the weight adjustment, the intercept matching algorithm needs to impute data that agree with the auxiliary margin, based on the new weights. Both the true standard deviations and average estimated standard deviations of $\alpha_1$ and $\beta_2$ are quite large. As when design weights for unit respondents are known, the average estimated standard deviations of $\pmb{\alpha}$ and $\pmb{\beta}$ (except for $\alpha_1$ and $\beta_2$) based on multiple imputation combining rules are quite close to the corresponding true standard deviations, because those estimates  (except for $\alpha_1$ and $\beta_2$) do not involve auxiliary margins directly. 

\begin{table}
\centering
\caption[Simulation results for parameters when $\alpha_1 = -0.5$, $\beta_1=0.4$, $\beta_2 = -0.5$.]{Simulation results for model parameters when using $N$ or $\hat{N}$ to generate imputations.  Simulations generated with $(\alpha_1, \beta_1, \ eta_2) =(-0.5,0.4, -0.5)$ and 50 multiple imputation data sets. The``Truth" column refers to the parameters used to generate the population.}
\begin{tabular}{cccccccc}
\hline
& \multicolumn{3}{c}{Estimate}  & \multicolumn{2}{c}{SD} & \multicolumn{2}{c}{Avg. Est. SD} \\
& Truth & $N$ & $\sum w_i^a$ & $N$ & $\sum w_i^a$ & $N$ & $\sum w_i^a$ \\
\hline
$\alpha_0$ & .300 &.291 & .291 & .016 & .017 & .018 & .018\\
$\alpha_1$ & -.500 & -.423 & -.416 & .218 & .243 & .246 & .241 \\
$\beta_0$ & .200 & .144 & .146 & .026 & .028 & .030 & .030\\
$\beta_1$ & .400 & .388 & .392 & .035 & .036 & .038 & .039\\
$\beta_2$ & -.500 & .021 & .010 & .225 & .267 & .291 & .316 \\
$\gamma_0$ & -1.05 & -1.06 & -1.06 & .024 & .024 & .026 & .026\\
$\gamma_1$ & .200 & .210 & .211 & .030 & .029 & .032 & .032\\
$T_{X2}$ &  79477 & 79508 & 79477 & 202.8 & 179.6 & 1858.5 & 1844.8 \\
$T_{X1}$ & 85981 & 85906 & 85961 & 198.2 & 189.8 & 1909.4 & 1906.6 \\
$X_2=0|X_1=0$ & .514 & .502 & .504 & .012 & .014 & .016 & .016 \\
$X_2=0|X_1=1$ & .344 & .350 & .349 & .007 & .007 & .013 & .013 \\
$X_1=0|X_2=0$ & .452 & .443 & .443 & .012 & .014 & .019 & .018 \\
$X_1=0|X_2=1$ & .290 & .298 & .296 & .007  & .008 & .011 & .011 \\
\hline
\end{tabular}
\label{pop1_N_W}
\end{table}

\begin{landscape}
\begin{table}
\footnotesize
\centering
\caption[Simulation results for totals and conditional margins assuming design weights for unit respondents are not known and $\alpha_1 = -0.5$.]{Simulation results for totals and conditional margins assuming design weights for unit respondents are not known and $\alpha_0 = 0.3, \alpha_1 = -0.5, \beta_0 = 0.2, \nu_0 = -1.2, \gamma_0 = -1.05, \gamma_1 = 0.2$. Results averaged over 1000 sampled data sets from the simulated populations, using Poisson sampling and 50 multiple imputation data sets.}
\label{alphaWeak}
\begin{tabular}{l|ccc|cc|ccc|cc|c}
\hline
 & \multicolumn{3}{c|}{Estimate} & \multicolumn{2}{c|}{CI Coverage \% } & \multicolumn{3}{c|}{SD} & \multicolumn{2}{c|}{Avg. Est. SD} & 
$\sqrt{b_m/m}$ \\
 & Truth & MDAM & ICIN & MDAM & ICIN & Pre-SD & MDAM & ICIN & MDAM & ICIN & MDAM\\
 \hline
 \multicolumn{12}{l}{\textbf{(a)} $\pmb{\alpha_1 = -0.5, \beta_1 = 0.4, \beta_2 = -0.5}$} \\
\hspace{4mm}$T_{X_2}$ &  79477 & 79477 & 81986 & 100  & 55.2  & 1339.2 & 179.6 & 1357.2 & 1844.8  & 1320.0 & 184.8 \\
\hspace{4mm}$T_{X_1}$ & 85981 & 85961 & 86717 & 100  & 92.5  & 1384.0 & 189.8 & 1380.1 & 1906.6 & 1388.0 & 193.3 \\
\hspace{4mm}$X_2=0|X_1=0$ & .514 & .504 & .484 & 93  & 34.5  & .011 & .014 & .011 & .016 & .012 & .002 \\
\hspace{4mm}$X_2=0|X_1=1$ & .344 & .349 & .332 & 99.8  &  70.4  & .008 & .007 & .008 & .013 & .008 & .002 \\
\hspace{4mm}$X_1=0|X_2=0$ & .452 & .443 & .440 & 97.9  & 83.9  & .010 & .014 & .011 & .018 & .012 & .002 \\
\hspace{4mm}$X_1=0|X_2=1$ & .290 & .296 & .293 & 98.6  & 96.4  & .008 & .008  & .008 & .011 & .009 & .001 \\
\multicolumn{12}{l}{\textbf{(b)} $\pmb{\alpha_1 = -0.5, \beta_1 = 0.4, \beta_2 = -2.0}$} \\
\hspace{4mm}$T_{X_2}$ &  79499 & 79489 & 81227 & 100  & 75  & 1306.3  & 184.9 & 1349.5 & 1837.6 & 1316.9 & 183.4 \\
\hspace{4mm}$T_{X_1}$ & 79834 & 79817 & 82176 & 100  & 59  & 1233.9 & 177.3 & 1330.3 & 1764.6 & 1320.4 & 176.3 \\
\hspace{4mm}$X_2=0|X_1=0$ & .515 & .505 & .492 & 97.1  & 52.9  & .010 & .012 & .011 & .017 & .012 & .002 \\
\hspace{4mm}$X_2=0|X_1=1$ & .329 & .336 & .328 & 99.8  & 96.1  & .008 & .007 & .007 & .012 & .008 & .001 \\
\hspace{4mm}$X_1=0|X_2=0$ & .512 & .502 & .483 & 96.7  & 28.7 & .010 & .012 & .010 & .016 & .012 & .002 \\
\hspace{4mm}$X_1=0|X_2=1$ & .327 & .333 & .320 & 99.6  & 91.9  & .008 & .007 & .008 & .012 & .009 & .001 \\
\multicolumn{12}{l}{\textbf{(c)} $\pmb{\alpha_1 = -0.5, \beta_1 = 2.0, \beta_2 = -0.5}$} \\
\hspace{4mm}$T_{X_2}$ &  79026  & 79016 & 80115 & 100  & 89.2  & 1277.3 & 178.8 & 1281.4 & 1800.0 & 1288.1 & 179.7 \\
\hspace{4mm}$T_{X_1}$ & 107451 & 105915 & 108007 & 100  & 95.1  & 1470.4 & 272.5 & 1453.4 & 2056.0 & 1503.9 & 201.9 \\
\hspace{4mm}$X_2=0|X_1=0$ & .948 & .944 & .950 & 100  & 94.9  & .007 & .007 & .007 & .013 & .007 & .002 \\
\hspace{4mm}$X_2=0|X_1=1$ & .277 & .268 & .270 & 100  & 83.9  & .006 & .003 & .006 & .014 & .007 & .002 \\
\hspace{4mm}$X_1=0|X_2=0$ & .453 & .477 & .453 & 85.9  & 97.7  & .010 & .016 & .010 & .026 & .012 & .003 \\
\hspace{4mm}$X_1=0|X_2=1$ & .017 & .019 & .016 & 100  & 92.3  & .002 & .003 & .002 & .005 & .002 & $<.001$ \\
\multicolumn{12}{l}{\textbf{(d)} $\pmb{\alpha_1 = -0.5, \beta_1 = 2.0, \beta_2 = -2.0}$} \\
\hspace{4mm}$T_{X_2}$ &  79592 & 79572 & 77283 & 100  & 56.5  & 1344.3 & 176.4 & 1303.6 & 1727.2 & 1240.2 & 169.8 \\
\hspace{4mm}$T_{X_1}$ & 102095 & 101788 & 104067 & 100  & 73.9  & 1557.1 & 336.7 & 1538.9 & 1746.4 & 1450.0 & 145.0 \\
\hspace{4mm}$X_2=0|X_1=0$ & .874 & .900 & .949 & 95.8  & 0  & .009 & .013 & .007 & .031 & .008 & .004 \\
\hspace{4mm}$X_2=0|X_1=1$ & .259 & .250 & .272 & 88.6  & 53.8  & .006 & .006 & .006 & .008 & .007 & $<.001$ \\
\hspace{4mm}$X_1=0|X_2=0$ & .509 & .528 & .496 & 74.1  & 86.6  & .010 & .013 & .010 & .014 & .012 & .001 \\
\hspace{4mm}$X_1=0|X_2=1$ & .050 & .040 & .020 & 96.5  & 0  & .004 & .005 & .003 & .013 & .003 & .002 \\
\hline
\end{tabular}
\end{table}
\end{landscape}

\begin{landscape}
\begin{table}
\footnotesize
\centering
\caption[Simulation results for totals and conditional margins assuming design weights for unit respondents are not known and $\alpha_1 = -2.0$.]{Simulation results for totals and conditional margins assuming design weights for unit respondents are not known and $\alpha_0 = 0.3, \alpha_1 = -2.0, \beta_0 = 0.2, \nu_0 = -1.2, \gamma_0 = -1.05, \gamma_1 = 0.2$. Results averaged over 1000 data sets sampled from the simulated populations, using Poisson sampling and 50 multiple imputation data sets.}
\begin{tabular}{l|ccc|cc|ccc|cc|c}
\hline
 & \multicolumn{3}{c|}{Estimate} & \multicolumn{2}{c|}{CI Coverage \%} & \multicolumn{3}{c|}{SD} & \multicolumn{2}{c|}{Avg. Est. SD} & 
$\sqrt{b_m/m}$ \\
 & Truth & MDAM & ICIN & MDAM & ICIN & Pre-SD & MDAM & ICIN & MDAM & ICIN & MDAM \\
 \hline
 \multicolumn{12}{l}{\textbf{(a)} $\pmb{\alpha_1 = -2.0, \beta_1 = 0.4, \beta_2 = -0.5}$} \\
\hspace{4mm}$T_{X_2}$ &  73451 & 73685 & 81569 & 100  & 0  & 1274.3 & 383.2 & 1347.4 & 1641.1 & 1315.9 & 147.2 \\
\hspace{4mm}$T_{X_1}$ & 84985 & 84966 & 86193 & 100  & 86.6  & 1355.2 & 193.0 & 1370.1 & 1895.3 & 1376.5 & 192.8 \\
\hspace{4mm}$X_2=0|X_1=0$ & .581  & .548 & .492 & 55.3  & 0  & .010 & .017 & .011 & .017 & .012 & .002\\
\hspace{4mm}$X_2=0|X_1=1$ & .375  & .389 & .332 & 97.8  & 0  & .008 & .007 & .009 & .015 & .008 & .002\\
\hspace{4mm}$X_1=0|X_2=0$ & .469  & .445 & .448 & 89.9  & 58  & .009 & .017 & .010 & .023 & .011 & .003\\
\hspace{4mm}$X_1=0|X_2=1$ & .277 & .296 & .294 & 46.6  & 50.8   & .008 & .008 & .008 & .009 & .009 & .001\\
\multicolumn{12}{l}{\textbf{(b)} $\pmb{\alpha_1 = -2.0, \beta_1 = 0.4, \beta_2 = -2.0}$} \\
\hspace{4mm}$T_{X_2}$ & 73410 & 73534 & 80869 & 100  & 0  & 1286.1  & 275.4 & 1349.4 & 1695.4 & 1315.7 & 158.5 \\
\hspace{4mm}$T_{X_1}$ & 79506 & 79495 & 82028 & 100  & 54.6  & 1335.5 & 171.5 & 1363.5 & 1776.0 & 1319.1 & 178.9 \\
\hspace{4mm}$X_2=0|X_1=0$ & .617 & .549 & .491 & 1.9  & 0  & .010 & .016 & .011 & .016 & .012 & .002 \\
\hspace{4mm}$X_2=0|X_1=1$ & .336 & .380 & .333 & 1.6  & 94.2  & .008 & .007 & .008 & .014 & .008 & .002 \\
\hspace{4mm}$X_1=0|X_2=0$ & .555 & .495 & .480 & 14.2  & 0  & .010 & .016 & .011 & .021 & .012 & .003 \\
\hspace{4mm}$X_1=0|X_2=1$ & .281 & .330 & .323 & 0  & 0.2  & .008 & .008 & .008 & .010 & .009 & .001 \\
\multicolumn{12}{l}{\textbf{(c)} $\pmb{\alpha_1 = -2.0, \beta_1 = 2.0, \beta_2 = -0.5}$} \\
\hspace{4mm}$T_{X_2}$ &  73808 & 73804 & 78826 & 100  & 2.6  & 1292.5  & 173.2 & 1302.2 & 1724.2 & 1262.2 & 170.9 \\
\hspace{4mm}$T_{X_1}$ & 104446 & 103010 & 105932 & 100  & 83.6  & 1547.6 & 360.0 &  1548.0 & 2083.2 & 1473.7 & 212.1  \\
\hspace{4mm}$X_2=0|X_1=0$ & .965 & .955 & .950 & 84  & 52.7 & .005 & .007 & .007 & .007 & .007 & .001\\
\hspace{4mm}$X_2=0|X_1=1$ & .303 & .297 & .269 & 100  & 0  & .007 & .004 & .006 & .017 & .007 & .002 \\
\hspace{4mm}$X_1=0|X_2=0$ & .469 & .488 & .479 & 99  & 89.3  & .010 & .014 & .010 & .029 & .012 & .004 \\
\hspace{4mm}$X_1=0|X_2=1$ & .014 & .018 & .017 & 76.5  & 79.9  & .002 & .003 & .003 & .003 & .003 & $<.001$ \\
\multicolumn{12}{l}{\textbf{(d)} $\pmb{\alpha_1 = -2.0, \beta_1 = 2.0, \beta_2 = -2.0}$} \\
\hspace{4mm}$T_{X_2}$ &  73582 & 73572 & 75242 & 100  & 73.2  & 1283.7  & 163.3 & 1277.7 & 1661.6 & 1208.5 & 164.0 \\
\hspace{4mm}$T_{X_1}$ & 98814 & 97638 & 101173 & 100  & 61.9  & 1492.8 & 381.4 & 1507.6 & 1685.2 & 1415.7 & 138.7 \\
\hspace{4mm}$X_2=0|X_1=0$ & .961 & .942 & .949 & 77.5  & 69.8  & .005 & .007 & .007 & .012 & .007 & .001 \\
\hspace{4mm}$X_2=0|X_1=1$ & .269 & .268 & .272 & 100  & 94.5  & .007 & .004 & .006 & .012 & .007 & .001 \\
\hspace{4mm}$X_1=0|X_2=0$ & .556 & .563 & .526 & 99.2  & 26.9  & .010 & .010 & .010 & .017 & .012 & .002 \\
\hspace{4mm}$X_1=0|X_2=1$ & .018 & .028 & .022 & 70.9  & 87.6  & .002  & .003 & .003 & .006 & .003 & .001 \\
\hline
\end{tabular}
\label{alphaStrong}
\end{table}
\end{landscape}

\begin{table}
\centering
\caption[Simulation results of parameters when design weights for unit respondents are not known.]{Simulation results of parameters when design weights are not available in the survey.  The  column labelled ``Truth" includes the parameter estimates used to generate the population.  Results based on 50 multiple imputation data sets.}
\begin{tabular}{c|ccc|ccc|cc}
\hline
 & \multicolumn{3}{c|}{Estimate}  & \multicolumn{3}{c|}{SD} & \multicolumn{2}{c}{Avg. Est. SD} \\
 & Truth & MDAM & ICIN & Pre-SD & MDAM & ICIN & MDAM & ICIN \\
 \hline
$\alpha_0$ & .3 & .291 & .291 & .017 & .017 & .017 & .018 & .018 \\
$\alpha_1$ & -.5 & -.416 & .008 &.037 & .243 & .013 & .241 & .057 \\
$\beta_0$ & .2 & .146 & .146 & .027 & .028 & .028 & .030 & .030 \\
$\beta_1$ & .4 & .392 & .392 & .034 & .036 & .036 & .039 & .038 \\
$\beta_2$ & -.5 & .010 & .052 & .034 & .267 & .012 & .316 & .058 \\
$\gamma_0$ & -1.05 & -1.06 & -1.06 & .024 & .024 & .024 & .026 & .026 \\
$\gamma_1$ & .2 & .211 & .211 & .029 & .029 & .029 & .032 & .032 \\
\hline
\end{tabular}
\label{parameter}
\end{table}

\clearpage

\section{2018 CPS Model and Intercept Matching Algorithm}\label{CPSmodel}

After taking both steps of the MD-AM framework, we analyze the 2018 CPS data with the following model.  In writing the model, we suppress the variables in the conditioning to save space.  $I[\cdot]$ is the indicator function that equals one when the value inside the brackets is true and equals zero otherwise. 
\begin{align}
U_i & \sim Bernoulli(\pi_i^U),\, logit(\pi_i^U) = \nu_0 \label{CPS_1}\\
S_i & \sim Bernoulli(\pi_i^G),\, logit(\pi_i^G) = \alpha_0^G + \alpha_1^G U_i \label{CPS_S}\\
E_i & \sim Multinomial(\mathbb{P}[E_i = e]),\, log(\mathbb{P}(E_i = e)/\mathbb{P}(E_i = 1)) =  \alpha_{0,e}^{E} + \alpha_{1,e}^{E}S_i + \alpha_{2,e}^{E}U_i \label{CPS_E}\\
C_i & \sim Multinomial(\mathbb{P}[C_i = c]),\, log(\mathbb{P}(C_i = c)/\mathbb{P}[C_i = 1]) = \alpha_{0,c}^{C} + \alpha_{1,c}^{C} S_i + \sum_{e=2}^4 \alpha_{2,e}^{C} I[E_i = e] \label{CPS_E}\\
 A_i & \sim Multinomial(\mathbb{P}[A_i = a]),\, log(\mathbb{P}[A_i = a]/\mathbb{P}[A_i = 1]) = \alpha_{0,a}^{A} + \alpha_{1,a}^{A} S_i \nonumber \\
 & \qquad + \sum_{e=2}^4 \alpha_{2,e,a}^A I[E_i = e] + \sum_{c=2}^3\alpha_{3,c,a}^A I[C_i = c] \label{CPS_A}\\
V_i & \sim Bernoulli(\pi_i^V),\, logit(\pi_i^V) = \alpha_{0}^{V} + \alpha_{1}^{V}S_i + \sum_{e=2}^4 \alpha_{2,e}^V I[E_i = e]  + \sum_{c=2}^3 \alpha_{3,c}^V I[C_i = c]  \nonumber\\
& \qquad + \sum_{a=2}^6 \alpha_{4,a}^V I[A_i = a] + \sum_{e=2}^4\alpha_{5,e}^V I[S_i = 1, E_i = e] + \sum_{a=2}^6\alpha_{6,a}^V I[S_i = 1, A_i = a] \nonumber \\
& \qquad + \sum_{c=2}^3 \alpha_{7,c}^V I[S_i = 1, C_i = c] + \alpha_8^V U_i \label{CPS_V}.\\
R_i^{E} & \sim Bernoulli(\pi_i^E),\, logit(\pi_i^E) = \gamma_{0}^{E} + \gamma_1^E S_i  + \sum_{c=2}^3\gamma_{2,c}^E I[C_i = c] + \sum_{a=2}^6\gamma_{3,a}^E I[A_i = a] + \gamma_{4}^E V_i \label{CPS_RE} \\
R_i^{C} & \sim Bernoulli(\pi_i^C),\, logit(\pi_i^C) = \gamma_{0}^{C} + \gamma_1^C S_i  + \sum_{e=2}^4 \gamma_{2,e}^C I[E_i = e] + \sum_{a=2}^6\gamma_{3,a}^C I[A_i = a] + \gamma_{4}^C V_i \label{CPS_RC} \\
R_i^{A} & \sim Bernoulli(\pi_i^A),\, logit(\pi_i^A) = \gamma_{0}^{A} + \gamma_1^A S_i  + \sum_{e=2}^4\gamma_{2,e}^A I[E_i = e] + \sum_{c=2}^3\gamma_{3,c}^A I[C_i=c]\label{CPS_RA} \\
P(R_i^{V} =1 \mid R_i^A = 1) & = 1 \\
R_i^{V}|R_i^A = 0  & \sim Bernoulli(\pi_i^V),\, logit(\pi_i^V) = \gamma_{0}^{V} + \gamma_1^V S_i  + \sum_{a=2}^6\gamma_{2,a}^V I[A_i = a] + \sum_{e=2}^4\gamma_{3,e}^V I[E_i = e] \nonumber \\
& + \sum_{c=2}^3\gamma_{4,e}^V I[C_i = c] \label{CPS_RV}.
\end{align}

The intercept matching algorithm for this model proceeds sequentially following the strategy described in the main text.  To give a sense of the updates, here we present the steps involving imputation of vote, which has a margin, and imputation of education, which does not have a margin.

Let $n_U$ be the number of unit nonrespondents. Starting at any iteration $t$, the steps to impute vote are as follows.
\begin{enumerate}
    \item Draw a value $\hat{T}_{V}^{(t+1)} \sim N(T_{V},{V}_{V})$.
\item Calculate the number of times $V_i^{(t+1)}$ should be imputed as 1 when $U_i = 1$ so that the weighted sum of $V^{(t+1)}$ is as close to $\hat{T}_{V}^{(t+1)}$ as possible; denote this number as $n_{V}$. Specifically,
\begin{equation}
n_{V} = \lfloor \frac{\hat{T}_{V}^{(t+1)} -\sum_{i \in \mathcal{D}} w_iV_{i}^{(t+1)}I(U_i = 0)}{\sum_{i \in \mathcal{D}} w_i I(U_i = 1)/n_U} \rfloor.
\end{equation}
\item Let $\hat{\alpha}^V$ and $V_{\hat{\alpha}^V}$ denote the maximum likelihood estimate and the variance of $\alpha^V = (\alpha_{0}^{V},\dots,\alpha_{7}^{V})$ based on the $U_i = 0$ observations. Sample ${\alpha^V}^{(t+1)}$ from its approximate posterior distribution, $N(\hat{\alpha}^V,V_{\hat{\alpha}^V})$.
\item Calculate the proportion of unit nonrespondents that should be imputed as 1 for $V$. Denote this number as $\hat{p_V} = \frac{n_{V}}{n_U}$. Then, calculate ${\alpha_8^V}^{(t+1)}$ so that 
\begin{eqnarray}
 logit(\hat{p}_V) &=& \mathbb{E}\Big[{\alpha_{0}^{V}}^{(t+1)} + {\alpha_{1}^{V}}^{(t+1)}S + \sum_{e=2}^4 {\alpha_{2,e}^V}^{(t+1)} I[E = e]  + \sum_{c=2}^3 {\alpha_{3,c}^V}^{(t+1)} I[C = c] \nonumber\\ 
 &+& \sum_{a=2}^6  {\alpha_{4,a}^V}^{(t+1)} I[A = a] 
+ \sum_{e=2}^4{\alpha_{5,e}^V}^{(t+1)} I[S = 1, E = e] + \sum_{a=2}^6{\alpha_{6,a}^V}^{(t+1)} I[S = 1, A = a] \nonumber \\
&+& \sum_{c=2}^3 {\alpha_{7,c}^V}^{(t+1)} I[S = 1, C = c] 
+ {\alpha_8^V}^{(t+1)}\Big].
\end{eqnarray}
\item Draw imputations of $V$ for unit nonrespondents, $V|U=1 \sim Bernoulli(1/d_v)$ where
\begin{eqnarray}
d_v &=& \big(1 + \exp(-{\alpha_{0}^{V}}^{(t+1)} - {\alpha_{1}^{V}}^{(t+1)}S - \sum_{e=2}^4 {\alpha_{2,e}^V}^{(t+1)} I[E = e]  - \sum_{c=2}^3 {\alpha_{3,c}^V}^{(t+1)} I[C = c] \nonumber \\
&-& \sum_{a=2}^6 {\alpha_{4,a}^V}^{(t+1)} I[A = a] - \sum_{e=2}^4{\alpha_{5,e}^V}^{(t+1)} I[S = 1, E = e] - \sum_{a=2}^6{\alpha_{6,a}^V}^{(t+1)} I[S = 1, A = a] \nonumber \\
&-& \sum_{c=2}^3 {\alpha_{7,c}^V}^{(t+1)} I[S = 1, C = c] - {\alpha_8^V}^{(t+1)})\big).
\end{eqnarray}
\item Draw imputations of $V^{(t+1)}$ when $(R_{i}^V = 1, U_i = 0)$ from its posterior distribution implied by \eqref{CPS_V}--\eqref{CPS_RC}.  We have 
\begin{align}
    p(V|R^V=1,U=0,\dots) \propto p(V|{\alpha^V}^{(t+1)},\dots) p(R^E|V,\dots) p(R^C|V,\dots)
\end{align}
\end{enumerate}

The steps to impute education are as follows.
\begin{enumerate}
    \item Let $\hat{\alpha}^C$ and $V_{\hat{\alpha}^C}$ denote the maximum likelihood estimate and the variance of $\alpha^C = (\alpha_{0}^{C},\dots,\alpha_{4}^{C})$ based on the $U_i = 0$ observations. Sample ${\alpha^C}^{(t+1)}$ from its approximate posterior distribution, $N(\hat{\alpha}^C,V_{\hat{\alpha}^C})$.
    \item Draw imputations of $C$ for unit nonrespondents: $C|U=1  \sim Multinom(\mathbb{P}[C = c]),\, log(\mathbb{P}(C = c)/\mathbb{P}[C = 1]) = {\alpha_{0,c}^{C}}^{(t+1)} + {\alpha_{1,c}^{C}}^{(t+1)} S + \sum_{e=2}^4 {\alpha_{2,e}^{C}}^{(t+1)} I[E = e]$.
    \item Draw imputations of $C^{(t+1)}$ when $(R_i^C=1,U_i=0)$ from its posterior distribution implied by \eqref{CPS_E}--\eqref{CPS_V}, \eqref{CPS_RE},\eqref{CPS_RA},\eqref{CPS_RV}.  Writing only the quantities tied to $C$ in the conditioning to save space, we have 
    \begin{align}
        p(C|R^C=1,U=0,\dots) & \propto p(C|{\alpha^C}^{(t+1)},\dots) p(A|C,\dots) p(V| C, \dots) \nonumber \\
        & \quad \times p(R^E|C,\dots) p(R^A|C,\dots) p(R^V|C, \dots)  
    \end{align}
\end{enumerate}

\section{Results from Measurement Error Modeling}\label{CPSmeasurementerror}

In the main text, we present results of a sensitivity analysis where we allow reporting errors among people who say they voted.  This requires modifying the intercept matching algorithm.  Specifically, we need to impute true vote $V_i$ for unit respondents with reported vote $Z_i=1$. This requires the following additional updates.  
\begin{enumerate}
    \item Sample 
    \begin{align}
        {\theta_1}^{(t+1)} &\sim Beta\Big(a_1 + \sum_{i \in \mathcal{D}} I(Z_i=1, V_i=0,C_i=1), b_1 + \sum_{i \in \mathcal{D}} I(Z_i=0, V_i=0,C_i=1) \Big), \\
    {\theta_2}^{(t+1)} &\sim Beta\Big(a_2 + \sum_{i \in \mathcal{D}} I(Z_i=1, V_i=0,C_i=2), b_2 + \sum_{i \in \mathcal{D}} I(Z_i=0, V_i=0,C_i=2)\Big), \\
    {\theta_3}^{(t+1)} &\sim Beta\Big(a_3 + \sum_{i \in \mathcal{D}} I(Z_i=1, V_i=0,C_i=3), b_3 + \sum_{i \in \mathcal{D}} I(Z_i=0, V_i=0,C_i=3) \Big).
    \end{align}
    \item Let $\hat{\alpha}^V$ and $V_{\hat{\alpha}^V}$ denote the maximum likelihood estimate and the variance of $\alpha^V = (\alpha_{0}^{V},\dots,\alpha_{7}^{V})$ based on the $U_i = 0, R^V_i = 0$ observations. Sample ${\alpha^V}^{(t+1)}$ from its approximate posterior distribution, $N(\hat{\alpha}^V,V_{\hat{\alpha}^V})$.
    \item Using $\theta_c^{(t+1)}$ and \eqref{CPS_V}, draw $V_i$ for $U_i=0,R^V_i=0$ units from
    \begin{align}
        p(V|Z,{\alpha^V}^{(t+1)},C=c,\dots) & \propto p(Z|V,{\alpha^V}^{(t+1)},C=c) p(V|{\alpha^V}^{(t+1)},\dots) \nonumber\\
        & = {\theta_c}^{(t+1)}  p(V|{\alpha^V}^{(t+1)},\dots).\label{Vmeaserr}
    \end{align}
\end{enumerate}
We do not include contributions to the likelihood in \eqref{Vmeaserr} from other variables, including $R^V$.  In our evaluations, we find that $Z$ and the model for vote provide most of the information about $V_i$, particularly since the model for $R^V$ is an ICIN model.

Results for the measurement error modeling with the 2018 CPS data are presented in Table \ref{marginal_CPS_Int_MDAM}, Table \ref{one_sub_Int_MDAM}, and Table \ref{two_sub_Int_MDAM}.  The two models overall yield similar estimates.

\begin{table}
\centering
\caption[Results from the CPS data analysis for marginal distributions.]{Estimated marginal distributions of sex, age, and race based on 50 imputations generated from the MD-AM hybrid missingness model with and without measurement error.  The auxiliary marginal percentage are is .52 for female, .699 for White, .218 for Black, .039 for Hispanic, and .044 for the remaining people. The $\sqrt{b_L/L}$ tends to estimate the multiple imputation variance more accurately for totals with known margins.}
\begin{tabular}{lrrrrrrr}
\hline
& \multicolumn{3}{c}{No Meas. Error} & & \multicolumn{3}{c}{With Meas. Error} \\
\hline
 & Prop & SD & $\sqrt{b_L/L}$  && Prop & SD &  $\sqrt{b_L/L}$\\
\hline
Male & .478 & .017 & $.002$ & & .474 & .014 & $<.001$ \\
Female & .522 & .017 &  $.002$ & & .526 & .014 & $<.001$ \\
\hline
White &  .699 & .013 & $.001$ & & .697 & .014 & .002 \\
Black &  .219 & .013 & $.001$ & & .220 & .012 & .001 \\
Hispanic & .038 & .005 & $<.001$ & & .039 & .005 & $<.001$ \\
Rest & .044 & .005 & $<.001$ & & .044 & .006 & $<.001$ \\
\hline
$(0,29]$ & .213 & .014 &  & & .214 & .017 &  \\
$(29,39]$ & .160 & .021 &  & & .158 & .019 &  \\
$(39,49]$ & .189 & .018 &  & & .190 & .019 &  \\
$(49,59]$ & .162 & .013 &  & & .162 & .014 &  \\
$(59,69]$ & .147 & .010 &  & & .149 & .011 &  \\
$(69,79]$ & .092 & .009 &  & & .092 & .009 &  \\
$>79$ & .036 & .005 &  & & .035 &  .004 &  \\
\hline
HS- & .372 & .012 &  & & .373 & .011 &  \\
Some College & .297 & .011 &  & & .297 & .011 &  \\
BA+ & .331 & .012 & & & .330 & .011 & \\
\hline
\end{tabular}
\label{marginal_CPS_Int_MDAM}
\end{table}

\begin{table}[h!]
\centering
\caption[Proportion of ``voted" in each subgroup of the CPS data.]{Estimated proportion who voted in each subgroup of the CPS data based on 50 imputations generated from the hybrid missingness MD-AM model with and without measurement error. Results based on 50 multiple-imputations for each model.  The auxiliary margin for voted in North Carolina is .49.  $\sqrt{b_L/L}$ for proportion who voted is $.002$.}
\begin{tabular}{lrrrrrr}
\hline
& \multicolumn{2}{c}{No Meas. Error} & & \multicolumn{2}{c}{With Meas. Error} \\
\hline
 & Prop & SD  & & Prop & SD  \\
\hline
Full & .501 & .016 & & .501 & .016\\
Male & .499 & .021  & & .499 & .022\\
Female & .504 & .020 & & .503 & .020\\
\hline
White &  .510 & .018 & & .513 & .019\\
  Black &  .519 & .032  & & .520 & .037\\
Hispanic & .390 & .061  & & .364 & .065\\
Rest & .378 & .057  & & .346 & .055\\
\hline
$(0,29]$ & .325 & .027  & & .296 & .027\\
$(29,39]$ & .398 & .038  & & .393 & .036\\
  $(39,49]$ & .492 & .041  & & .508 & .038\\
$(49,59]$ & .595 & .038  & & .587 & .032\\
 $(59,69]$ & .700 & .035  & & .717 & .032\\
 $(69,79]$ & .672 & .039  &  & .681 & .040\\
$>79$ & .467 & .060 & & .454 & .061\\
\hline
HS- &  .347 & .020 &  &.343 & .021\\
Some College & .519 & .023 & & .514 & .026\\
 BA+ & .659 & .029  & & .669 & .024\\
\hline
\end{tabular}
\label{one_sub_Int_MDAM}
\end{table}

\begin{table}[h!]
\centering
\caption[Proportion of ``voted" in each subgroups of gender crossed with race/age of the CPS data.]{Estimated proportion who voted in subgroups defined by sex crossed with race/age/education fir the 2018 CPS data based on the hybrid missingness MD-AM model with and without measurement error. Results based on  50 multiple imputations with each model.}
\begin{tabular}{lrrrrrr}
\hline
& \multicolumn{2}{c}{No Meas. Error} & & \multicolumn{2}{c}{With Meas. Error} \\
\hline
 & Prop & SD & & Prop & SD \\
\hline
Male, White &  .513 & .023 & & .516 & .025 \\
Male, Black &  .475 & .047  & & .473 & .050 \\
Male, Hispanic & .459 & .100  & & .456 & .094  \\
Male, Rest & .427 & .081  & & .401 & .082 \\
Female, White &  .508 & .023 & & .510 & .024 \\
 Female, Black &  .557 & .038  & & .559 & .044 \\
Female, Hispanic & .323 & .080  & & .277 & .090 \\
Female, Rest & .334 & .072  & & .295 & .071 \\
\hline
Male, $(0,29]$ & .313 & .038  & & .291 & .038 \\
Male, $(29,39]$ & .396 & .049  & & .397 & .049 \\
 Male, $(39,49]$ & .451 & .053  & & .463 & .053\\
Male, $(49,59]$ & .577 & .048  & & .575 & .045 \\
 Male, $(59,69]$ & .714 & .049  & & .729 & .044\\
 Male, $(69,79]$ & .764 & .050  & & .775 & .054\\
Male, $>79$ & .460 & .094  & & .431 & .096\\
Female, $(0,29]$ & .337 & .038 &  & .300 & .038\\
Female, $(29,39]$ & .401 & .050  & & .389 & .046\\
 Female, $(39,49]$ & .530 & .048 & & .546 & .046\\
 Female, $(49,59]$ & .594 & .046  & & .598 & .044\\
 Female, $(59,69]$ & .687 & .043  & & .706 & .046\\
 Female, $(69,79]$ & .596 & .054  & & .603 & .051\\
Female, $>79$ & .473 & .072  &  & .467 & .071\\
\hline
Male, HS- & .347 & .026 & & .341 & .030 \\
Male, Some College & .520 & .035 &  & .513 & .040\\
 Male, BA+ & .677 & .039  & & .696 & .035\\
Female, HS- &  .348 & .026  & & .345 & .029\\
Female, Some College & .519 & .031 &  & .514 & .034\\
Female, BA+ & .645 & .034 & & .648 & .031\\
\hline
White, HS- & .341 & .023 && .339 & .025\\
White, Some College & .512 & .026 && .511 & .030\\
White, BA+ & .663 & .030 && .675 & .026 \\
Male, White, HS- & .341 & .031 && .339 & .033\\ 
\hline
\end{tabular}
\label{two_sub_Int_MDAM}
\end{table}

\bibliographystyle{natbib}
\bibliography{lit}